\documentclass[twocolumn]{aastex7}

\usepackage{amsmath}
\usepackage{subcaption}

\newcommand{\revisiongraphic}[2][]{%
\IfFileExists{#2}{\includegraphics[#1]{#2}}{%
\fbox{\parbox[c][32mm][c]{0.92\linewidth}{\centering
Figure pending regeneration.\\\texttt{\detokenize{#2}}}}}}

\begin{document}

\title{\textsc{GPUPhot}: Kernel-Based Algorithms for Point-Source Detection and\\
Photometry with a Spatially Variable PSF}

\author[orcid=0000-0002-8134-2592,gname=Miguel R.,sname=Alarcon]{Miguel R. Alarcon}
\affiliation{Light Bridges, Observatorio Astronómico del Teide. Carretera del Observatorio del Teide, s/n, Güímar, Canarias, Spain}
\affiliation{Instituto de Astrofísica de Canarias (IAC), C/ Vía Láctea, s/n, E-38205, La Laguna, Spain}
\affiliation{Departamento de Astrofísica, Universidad de La Laguna (ULL), E-38206 La Laguna, Canarias, Spain}
\email[show]{mra@mralarcon.com}

\author[orcid=0000-0003-1044-154X,gname=Samuel,sname=Lemes-Perera]{Samuel Lemes-Perera}
\affiliation{Light Bridges, Observatorio Astronómico del Teide. Carretera del Observatorio del Teide, s/n, Güímar, Canarias, Spain}
\affiliation{Departamento de Ingeniería Informática y de Sistemas, Universidad de La Laguna (ULL), E-38206 La Laguna, Canarias, Spain}

\email{samuel@lightbridges.es}

\author[orcid=0000-0002-2394-0711,gname=Miquel,sname=Serra-Ricart]{Miquel Serra-Ricart}
\affiliation{Light Bridges, Observatorio Astronómico del Teide. Carretera del Observatorio del Teide, s/n, Güímar, Canarias, Spain}
\affiliation{Instituto de Astrofísica de Canarias (IAC), C/ Vía Láctea, s/n, E-38205, La Laguna, Spain}
\affiliation{Departamento de Astrofísica, Universidad de La Laguna (ULL), E-38206 La Laguna, Canarias, Spain}
\email{miquel@lightbridges.es}

\author[orcid=0000-0002-9214-337X,gname=Javier,sname=Licandro]{Javier Licandro}
\affiliation{Instituto de Astrofísica de Canarias (IAC), C/ Vía Láctea, s/n, E-38205, La Laguna, Spain}
\affiliation{Departamento de Astrofísica, Universidad de La Laguna (ULL), E-38206 La Laguna, Canarias, Spain}
\email{jlicandr@iac.es}

\begin{abstract}
Modern imaging surveys generate data at rates that exceed the throughput of photometry pipelines built on iterative, per-source fitting, a construct that does not map onto parallel hardware. We present a set of algorithms for point-source detection and photometry reformulated so that the entire reduction runs natively on a Graphics Processing Unit (GPU), implemented in \textsc{GPUPhot}. Their design rests on a single principle: every stage is cast as a convolution with a fixed kernel or an element-wise operation on full-frame maps. The algorithms are designed to operate on wide-field images with a spatially variable Point Spread Function (PSF), where the field-dependent PSF is described through a small set of fixed, flux-conserving principal-component eigenmodes with position-dependent coefficients, so that the spatially varying matched filter decomposes exactly into a few genuine convolutions while retaining its maximum-likelihood foundation. We show that, even in fields with strong background gradients and markedly inhomogeneous PSFs, the framework achieves sub-pixel astrometric accuracy and few-millimagnitude photometric precision for bright sources. The pipeline is already in production at the telescopes of the Teide Observatory, where it has processed 3.7 million images since 2023, amounting to 1.2~PB of data, at effective end-to-end rates below 15~s per 150-Mpx image, from ingestion of the raw frame to a source catalog with astrometric and photometric solutions. Because the same primitives underlie every stage, the individual algorithms are self-contained and applicable well beyond photometry. They form the basis of a complete, openly developed software package, released publicly with this paper.
\end{abstract}


\received{September 6, 2026}
\revised{September 17, 2026}
\accepted{October 8, 2026}

\section{Introduction}
\label{sec:intro}

Wide-field imaging surveys have become the primary engine of discovery in modern optical astronomy. Facilities such as Pan-STARRS \citep{Chambers2016}, ATLAS \citep{Tonry2018}, ASAS-SN \citep{Kochanek2017}, Tomo-e Gozen \citep{Sako2018}, the Zwicky Transient Facility \citep[ZTF;][]{Bellm2019}, the Dark Energy Camera \citep[DECam;][]{Flaugher2015}, the Wide Field Survey Telescope \citep[WFST;][]{Cai2025} and, most recently, the Vera C.\ Rubin Observatory Legacy Survey of Space and Time \citep[LSST;][]{IveziC2019} repeatedly image large areas of the sky at high cadence, turning the static view of the sky into a dynamic, time-resolved one. This shift has been accompanied by a steep growth in data rates: a single modern mosaic camera delivers hundreds of megapixels per exposure, and a single night can generate from hundreds of gigabytes to terabytes of imaging that must be reduced, increasingly, in near real time, if time-critical phenomena are to be identified while they remain observable. The reduction software, rather than the telescope, is now frequently the bottleneck that sets the scientific throughput of a survey.

At the heart of essentially every such pipeline lies the detection and photometry of point sources, and the quantity that governs both is the point spread function (PSF), the response of the atmosphere--telescope--detector system to a point of light. Accurate astrometry and photometry demands a faithful model of the PSF. Over four decades a rich toolbox has been developed for the purpose: aperture photometry, including its optimal, growth-curve and critically-sampled variants \citep{Stetson1990, Lauer1999, Bickerton2013}; PSF-fitting photometry, from the classical DAOPHOT \citep{Stetson1987} and DoPHOT \citep{Schechter1993} to discrete-PSF formulations \citep{Mighell2005}; integrated detection-and-measurement frameworks such as SExtractor with PSFEx \citep{Bertin2011} and the Hyper Suprime-Cam/LSST science pipeline \citep{Bosch2018}; and, more recently, machine-learning approaches \citep{Yang2023, Winecki2024}. A difficulty common to all of them is that, across a wide field, the PSF is not constant: optical aberrations, detector tilt and atmospheric seeing gradients make it vary systematically over the focal plane. Capturing this variation is essential for both unbiased detection and precise photometry, and has been addressed either by expanding the PSF on a spatial polynomial basis \citep{Bertin2011, Piotrowski2013} or by describing its variation through a principal-component decomposition \citep{Jarvis2004, Bai2025}.

Most of these methods were designed when data volumes were small enough that human supervision, rather than computational throughput, set the pace of the analysis; they typically rely on iterative, per-source least-squares fitting that does not scale to the data rates quoted above. The natural way to close the gap is massively parallel hardware. Graphics Processing Units (GPUs) apply the same operation to many data elements at once and are exceptionally well suited to the array arithmetic and Fast Fourier Transforms (FFTs) that underlie image reduction. Several recent efforts have exploited this: \citet{Niwano2021} obtained a fortyfold speed-up of a full reduction pipeline by porting it to the GPU with CuPy; \citet{Cao2025} assembled a GPU-accelerated preprocessing framework for the GWAC survey; and \citet{Hu2026} built a real-time DECam transient-detection pipeline around SFFT, a GPU-accelerated, Fourier-domain image-subtraction algorithm. These works demonstrate the value of GPU acceleration, but they tend either to port existing CPU algorithms step by step or to accelerate a single stage, most often image subtraction. What has been missing is a photometry pipeline conceived from the outset so that every stage reduces to the same GPU-friendly primitives, and in which the spatially variable PSF is handled within that framework rather than as an exception to it.

In this paper we present the algorithmic foundation of \textsc{GPUPhot}, a GPU-native software suite for point-source detection and photometry that fills this gap. Its single design principle is that every stage of the reduction is cast either as a convolution with a fixed kernel or as an element-wise operation on full-frame maps, so that the whole reduction runs natively on a GPU even where the underlying problem is not naturally a convolution.

\begin{figure*}[t!]
\centering
\includegraphics[width=\linewidth]{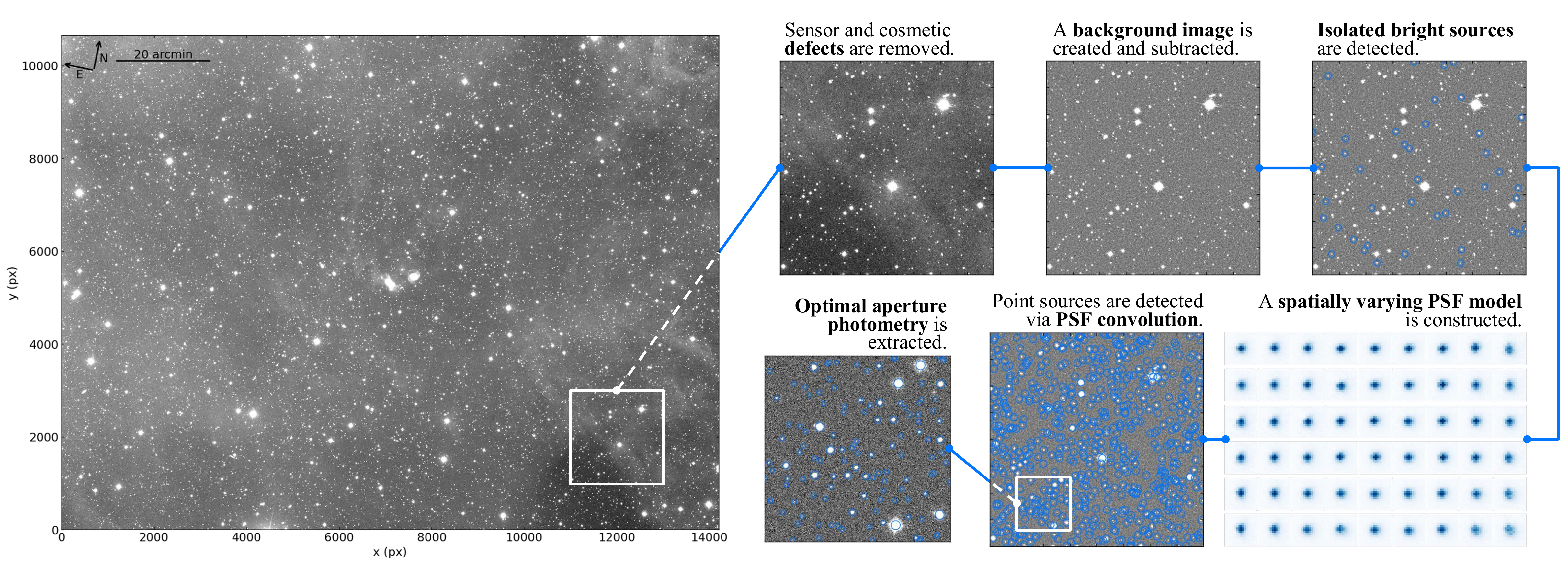}
\caption{Overview of the \textsc{GPUPhot} pipeline. Defect correction and background estimation precede isolated-star selection, spatially varying PSF modeling, matched-filter source detection and adaptive aperture photometry. Representative image stamps illustrate the successive stages.}
\label{fig:summary}
\end{figure*}

\textsc{GPUPhot} is already in routine operation as the real-time reduction pipeline of the Two-meter Twin Telescope (TTT\footnote{\url{ttt.iac.es}}), the Transient Survey Telescope (TST\footnote{\url{tst.iac.es}}) and ATLAS-Teide \citep{Licandro2023} at the Teide Observatory, where it processes the data streams of both their CCD and sCMOS cameras and has produced the photometry for a number of published studies \citep[e.g.][]{Reddy2024, Popescu2025, Durech2025, delaFuente2025, Rodriguez2026, Alarcon2026a, Alarcon2026b, Alarcon2026c}. With this paper we release \textsc{GPUPhot} publicly\footnote{\url{https://github.com/Light-Bridges/GPUPhot}} as the foundation of a complete, openly developed software suite, so that the wider community can adopt it for its own instruments and surveys.
The processing sequence is summarized in Figure~\ref{fig:summary}, the following sections describe its individual stages. Sections~\ref{sec:defects}--\ref{sec:photometry} describe the algorithms of each stage, covering defect correction and background estimation, the PCA-based PSF model, matched-filter detection and aperture photometry. Section~\ref{sec:results} validates the pipeline through a controlled injection--recovery experiment and against external catalogs, Section~\ref{sec:gpu} discusses the GPU implementation and its performance and Section~\ref{sec:conclusions} summarizes our conclusions.

\section{Sensor defect correction}\label{sec:defects}
\subsection{Cosmic rays} \label{sec:cr}

Cosmic rays are a common source of contamination in astronomical images, manifesting as high-intensity spikes that can severely compromise photometric precision. We have developed a GPU-optimized algorithm for their detection and mitigation, conceptually related to established methods such as L.A. Cosmic \citep{LACosmic}, which similarly relies on Laplacian edge detection and iterative filtering.

The algorithm begins by convolving the input image with a discrete Laplace operator, which is highly sensitive to the sharp intensity gradients characteristic of cosmic ray hits. Candidate pixels are flagged where the absolute deviation of the filtered image from its mean exceeds a threshold of $N_\sigma$ times the standard deviation of the Laplacian (default $N_\sigma = 3$), producing an initial binary mask. This mask is subsequently dilated with a $3\times3$ structuring element to ensure complete spatial coverage of each event. This fixed kernel size is motivated by the sub-pixel to few-pixel spatial extent of typical cosmic ray events on the detectors employed; a larger or parametric kernel would risk encroaching on the wings of point sources in well-sampled images.

\begin{figure*}[t!]
\centering
\begin{subfigure}[t]{0.485\textwidth}
\centering
\revisiongraphic[width=\linewidth]{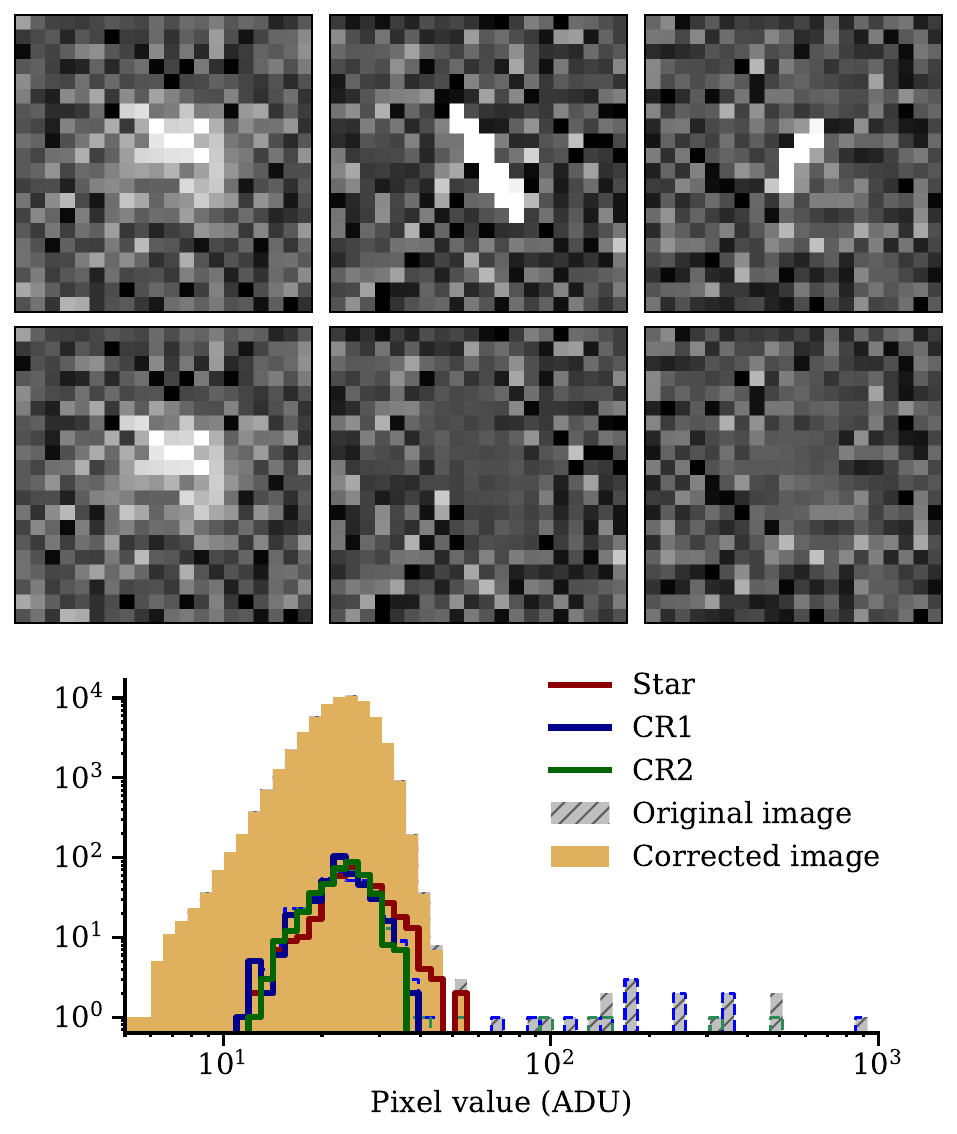}
\caption{Cosmic-ray correction.}
\label{fig:CR_filter}
\end{subfigure}\hfill
\begin{subfigure}[t]{0.485\textwidth}
\centering
\revisiongraphic[width=\linewidth]{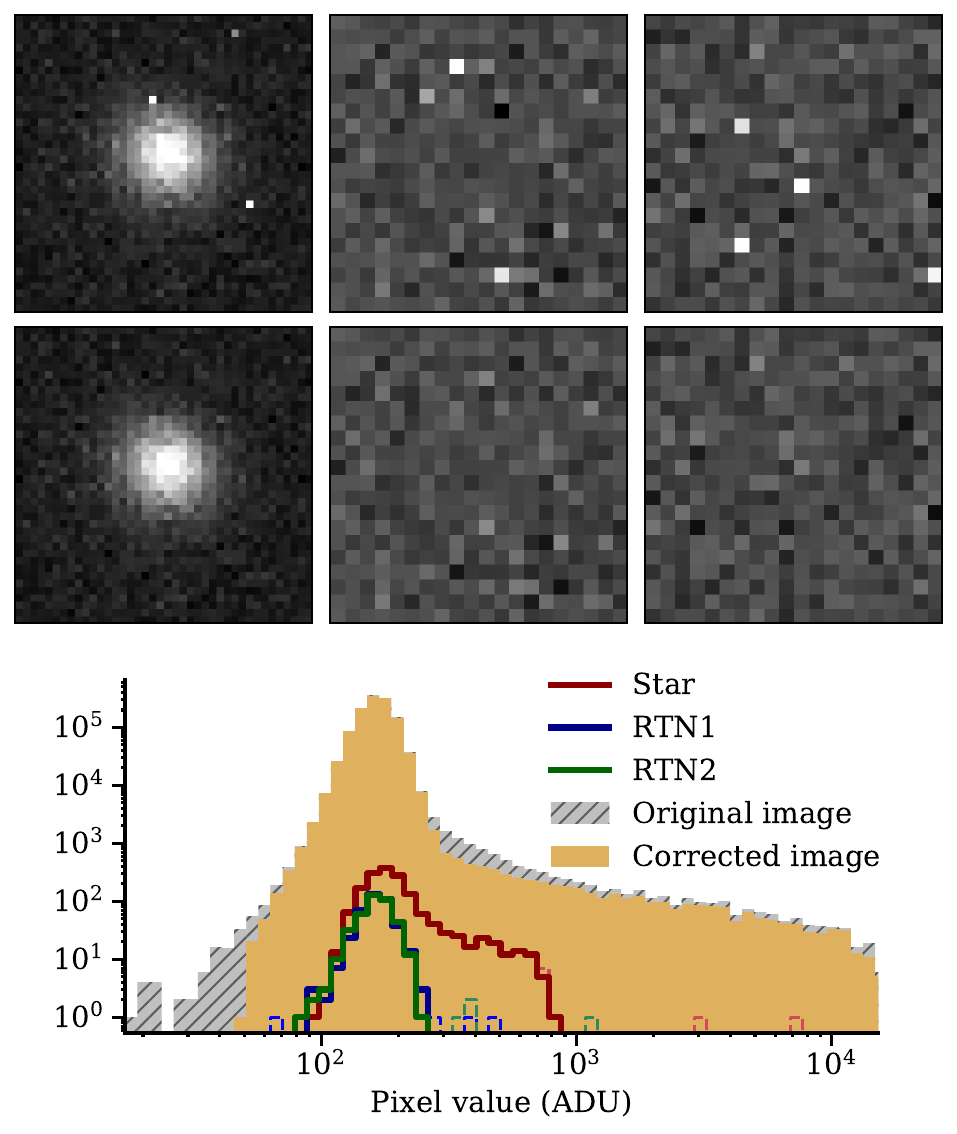}
\caption{RTN correction.}
\label{fig:SP_filter}
\end{subfigure}
\caption{Defect correction in real astronomical images. In each panel, the upper and middle rows show a stellar source and two affected regions before and after correction. The lower plot compares their pixel-value distributions: dashed outlines denote the original regional distributions and solid outlines the corrected distributions. Circle, square and triangle symbols identify the star and the two defects; hatching identifies only the original full-image distribution (gray), while the corrected distribution is unhatched. Panel (a) shows the two-stage cosmic-ray filter; panel (b) shows the isolated-pixel RTN filter.}
\label{fig:defects}
\end{figure*}

Pixels within the dilated mask are set to \texttt{NaN} in a working copy of the image, and an iterative Fast Fourier Transform (FFT)-based interpolation scheme (\texttt{fill\_nan\_fft}) reconstructs the affected regions by iteratively filling the masked pixels, repeating until no \texttt{NaN} values remain or the count of unfilled pixels is unchanged between successive iterations. The first reconstruction is used only to decide which candidate pixels should be corrected; it does not yet alter the original image. A candidate is corrected only if its reconstructed value lies below the global 95th percentile of the reconstructed image. These candidates are masked in the original image and filled by a second interpolation pass. All other pixels retain their original values. The motivation is that an isolated cosmic ray may interpolate to the background level, whereas the surrounding light of a star may keep its reconstructed core above the threshold. This distinction is only approximate: if the initial mask removes most of a stellar profile, the reconstructed core may also fall below the threshold and be incorrectly corrected. Figure~\ref{fig:CR_filter} illustrates the result for the example shown.

\subsection{Random Telegraph Noise} \label{sec:RTN}

Random Telegraph Noise (RTN), also referred to as salt-and-pepper noise in the image processing literature, arises from charge trapping and release events in individual pixels of solid-state detectors, producing sporadic positive or negative excursions that are uncorrelated with the local image structure. This phenomenon has been identified in some of the most widely used astronomical sensors, including the Sony~IMX455 and IMX411 sCMOS detectors \citep{Alarcon2023}. Unlike cosmic rays, which tend to affect contiguous pixel groups, RTN typically manifests as isolated single-pixel outliers, making it particularly amenable to identification through local statistical analysis.

Our mitigation algorithm proceeds in three stages. First, a median filter of kernel size $k$ (default $k=3$ px) is applied to produce a smooth reference image. The residual map, defined as the difference between the original and the median-filtered image, is then characterized statistically: its median $\mu_r$ and a robust estimate of its dispersion $\hat{\sigma}_r$ are computed, the latter derived from the Median Absolute Deviation (MAD) scaled to Gaussian units as
\begin{equation}
    \hat{\sigma}_r = 1.4826 \times \mathrm{MAD}(r - \mu_r),
\end{equation}
where $r$ denotes the residual map. Pixels are flagged as outliers if their residual satisfies
\begin{equation}
    r > \mu_r + N_+ \hat{\sigma}_r \quad \text{or} \quad
    r < \mu_r - N_- \hat{\sigma}_r,
\end{equation}
with default thresholds $N_+ = N_- = 5$, allowing for asymmetric detection if required.

In a second stage, a morphological isolation criterion is applied: flagged pixels are retained in the mask only if fewer than two of their eight immediate neighbors are also flagged. This condition suppresses the erroneous detection of structured image features--such as the wings of bright point sources or steep background gradients--which would otherwise produce clusters of adjacent outlier pixels. Finally, the value of each surviving masked pixel is replaced by the corresponding value from the median-filtered reference image, effectively reconstructing the affected pixel from its local neighborhood. This algorithm efficiently removes isolated single-pixel outliers  characteristic of RTN while leaving structured signals, including stellar profiles, unaffected, as illustrated in Figure~\ref{fig:SP_filter}.

\section{Background subtraction} \label{sec:background}

\begin{figure*}[ht!]
\centering
\includegraphics[width=\linewidth]{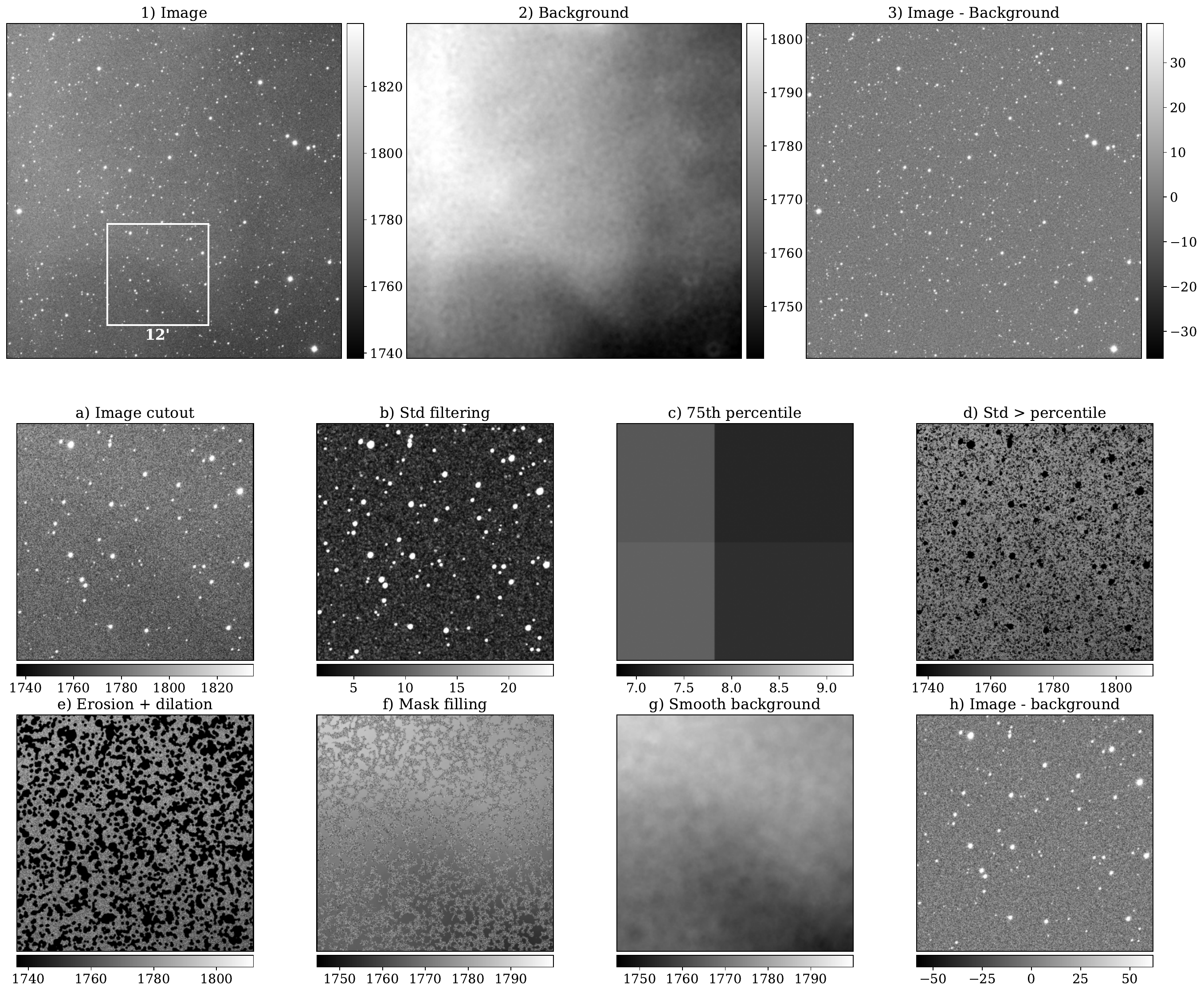}
\caption{Illustration of the background estimation algorithm applied to a real astronomical image. \textit{Top row:} full frame view of the original image (left), the estimated background map (center), and the background-subtracted image (right). The large-scale gradient across the field is effectively captured and removed. \textit{Bottom rows:} step-by step illustration of the algorithm on the subregion indicated by the white box in the top-left panel: (a) image cutout; (b) local standard deviation map after Gaussian smoothing; (c) coarse percentile threshold map derived from $T\times T$ tiles; (d) source mask from pixels exceeding the percentile threshold; (e) morphologically refined mask after erosion and dilation; (f) source-masked image after FFT-based interpolation; (g) final smoothed background map; (h) background-subtracted image cutout.
\label{fig:background}}
\end{figure*}

Accurate background estimation is a critical prerequisite for reliable source detection and photometry, particularly in crowded fields or images with spatially varying sky levels. Our algorithm produces a smooth, locally representative background map by combining tile-based percentile statistics with FFT-based convolution, and is designed to handle images of arbitrary size and pixel scale. The successive steps are illustrated in Figure~\ref{fig:background}.

The algorithm proceeds in four stages. First, the input image (Figure~\ref{fig:background}a) is convolved with a Gaussian kernel ($\sigma = 2$ px) to suppress pixel-level noise, and a local standard deviation map is computed over a sliding window of radius 2 arcsec (Figure~\ref{fig:background}b). This map serves as a proxy for the compactness of local emission: because the standard deviation is computed over a kernel smaller than the typical source size, point sources produce sharply elevated values compared to extended emission or smooth background gradients. This is a key distinction with respect to methods such as \citet{Akhlaghi2015}, which aim to separate all signal from the background, including extended emission. Our approach intentionally masks only compact point sources, preserving extended emission as part of the background to be estimated and subtracted.

In the second stage, the image is decomposed into non overlapping square tiles of size $T$ (default $T = 1000$ px), and the 75-th percentile of the local standard deviation is computed for each tile, discarding tiles containing \texttt{NaN} values. These per-tile values are reassigned to their corresponding regions to construct a coarse threshold map at the native image resolution (Figure~\ref{fig:background}c). Pixels whose local standard deviation exceeds this threshold are identified as source-dominated and masked (Figure~\ref{fig:background}d). The resulting mask is then morphologically refined through erosion and dilation with a structuring element of size $\max(k/s + 1,  2)$ px, where $s$ is the plate scale in arcsec\,px$^{-1}$ and $k$ is a smoothing scale in arcsec (default $k = 2$ arcsec), followed by a Gaussian smoothing of the mask boundaries ($\sigma = \max(k/s + 1,  3)$ px) with a threshold of 0.05 (Figure~\ref{fig:background}e).

In the third stage, masked pixels are set to \texttt{NaN} and an iterative FFT-based interpolation scheme (\texttt{fill\_nan\_fft}) reconstructs the source-masked regions using a filling aperture of $\max(r_f/s + 1,  5)$ px (default $r_f = 100$ arcsec), repeating until either no \texttt{NaN} values remain or the count of unfilled pixels is unchanged between successive iterations (Figure~\ref{fig:background}f). Finally, the reconstructed image is smoothed with an averaging kernel of radius $\max(r_a/s + 1,  3)$ px (default $r_a = 20$ arcsec) to produce the final background map (Figure~\ref{fig:background}g). The background-subtracted image (Figure~\ref{fig:background}h) is then used in all subsequent pipeline stages. FFT-based filling accommodates the contiguous holes left by cosmic-ray and source masking. For isolated RTN pixels, the median reference already computed for detection provides a local replacement without an additional filling pass.

The expected per-pixel variance is
\begin{equation}
    \sigma_x^2 = 
    \sigma_{B}^2 + \sigma_{R}^2 + \sigma_{D}^2,
    \label{eq:rms}
\end{equation}
where $x$ denotes the pixel position and all terms are expressed in e$^{-2}$. The first term, $\sigma_{B}^2$, is the Poisson variance of the sky background, equal to the local background level in electrons by virtue of Poisson statistics. The second, $\sigma_{R}^2$, is the readout noise variance, with $\sigma_{R}$ the per-pixel readout noise in e$^-$. The third, $\sigma_{D}^2 = D\,t$, is the dark current shot noise for a dark current rate $D$ in e$^-$\,px$^{-1}$\,s$^{-1}$ and exposure time $t$ in seconds; like $\sigma_{B}^2$, it is a Poisson variance whose value equals the number of dark electrons accumulated during the exposure. This noise map encodes the spatially varying detection limit imposed by Poisson fluctuations of the sky background, detector readout and dark current, and is used as the per-pixel significance threshold in the point-source detection stages.

\section{PSF Modeling}
\label{sec:psf}

Accurate photometry of point sources requires a faithful model of the Point Spread Function (PSF) at every position in the field. Optical aberrations, detector tilt and atmospheric seeing gradients cause the PSF to vary spatially across the focal plane, an effect that becomes significant in wide-field instruments where the field of view samples a substantial portion of the optical beam.

Following \citet{Zackay2017}, a background-subtracted image can be modeled as
\begin{equation}
    M = T * P + \epsilon,
    \label{eq:image_model_general}
\end{equation}
where $T$ is the true source flux distribution in $e^-$, $P$ is the PSF normalized to unit sum ($\sum_x P_x = 1$), $\epsilon$ is additive noise and $*$ denotes convolution. The per-pixel variance of $\epsilon$ is
\begin{equation}
    \mathrm{Var}[\epsilon_x] = \sigma_x^2 + 
    (T * P)_x,
    \label{eq:total_noise}
\end{equation}
combining the sky, read and dark current noise of Equation~(\ref{eq:rms}) with the Poisson variance of the source signal at pixel $x$. For an isolated point source of flux $\alpha$ at position $\mu$, the true sky contains a single non-zero pixel, $T_{x'} = \alpha \delta_{x',\mu}$, where $x'$ indexes the source plane and $\delta_{x',\mu}$ is the Kronecker delta.
Substituting into the convolution $(T * P)_x = \sum_{x'} T_{x'} P_{x-x'}$ collapses the sum to its single non-zero term $x' = \mu$, and Equation~(\ref{eq:image_model_general}) becomes
\begin{equation}
    M_x = \alpha P_{x-\mu} + \epsilon_x.
    \label{eq:point_source_model_general}
\end{equation}

\textsc{GPUPhot} constructs an empirical model of the PSF in three steps: detection and selection of isolated stars, construction of a position-independent reference PSF and characterization of spatial deviations from the reference via Principal Component Analysis (PCA).

\subsection{Isolated star detection}
\label{sec:psf/isolated}

PSF modeling requires a set of isolated, unsaturated point sources that adequately sample the spatial extent of the field. Because the PSF has not yet been determined at this stage of the pipeline, we approximate $P$ by a fixed Gaussian kernel $G$ of width $\sigma_G = 0.5$ arcsec, and Equation~(\ref{eq:point_source_model_general}) reduces to
\begin{equation}
    M_x = \alpha G_{x-\mu} + \epsilon_x.
    \label{eq:point_source_gaussian}
\end{equation}
To identify candidate sources we treat each discrete pixel-center position $\mu$ across the image as the hypothetical center of a point source and seek the statistic that best separates a real source from a background fluctuation, that is, the one that maximizes the detection signal-to-noise ratio at that position. Assuming background-dominated noise, with negligible source Poisson variance, and Gaussian noise uncorrelated between pixels, that flux is the maximum-likelihood estimate of $\alpha$ at fixed $\mu$ (see Appendix~\ref{app:detection} for a full derivation):
\begin{equation}
    \hat{\alpha}(\mu) = \frac{\sum_x M_x G_{x-\mu} / \sigma_x^2}{\sum_x G_{x-\mu}^2 / \sigma_x^2}.
    \label{eq:alpha_hat}
\end{equation}
The numerator cross-correlates the image with the kernel and the denominator normalizes it to units of flux, both weighting each pixel by its inverse variance $1/\sigma_x^2$ so that noisier pixels contribute less; $\hat{\alpha}(\mu)$ is thus the best-fitting flux of a source of profile $G$ centered at $\mu$. Normalizing this estimate by its own uncertainty yields the detection significance $\nu_G(\mu) = \hat{\alpha}(\mu)/\sigma_{\hat{\alpha}}(\mu)$, with $\sigma_{\hat{\alpha}}^2(\mu) = [\sum_x G_{x-\mu}^2/\sigma_x^2]^{-1}$, which is by construction the signal-to-noise ratio we sought. Sources are then identified as local maxima of $\nu_G$, where the point-source hypothesis is most strongly favored over background noise.

\begin{figure}[t!]
\centering
\revisiongraphic[width=\linewidth]{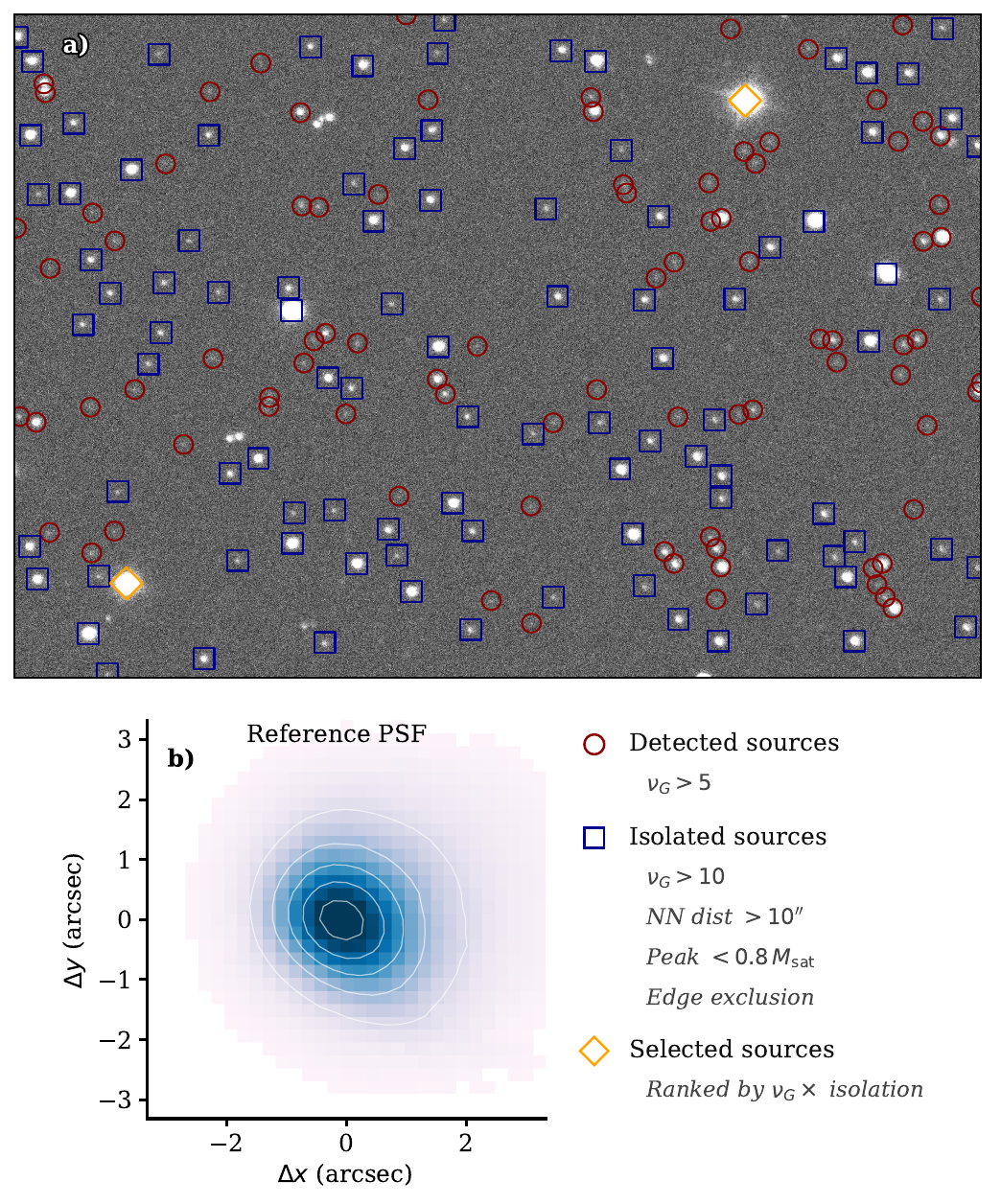}
\caption{Selection of isolated stars for PSF modeling, illustrated on a representative real astronomical image. \textbf{a)} background-subtracted cutout image stamp with detected sources overlaid. Red circles mark all initial detections with $\nu_G > 5$; blue squares mark candidates that survive the nearest-neighbor distance and saturation criteria; yellow diamonds mark the subset selected for PSF modeling after the central-field, isolation and brightness-ranking cuts. Only three of the 21 stars retained for the full field are visible in this 
cutout. \textbf{b)} reference PSF $P_0$ obtained by stacking the 21 selected normalized stars with iterative sigma-clipping.
\label{fig:psf_isolated}}
\end{figure}

The exact evaluation of $\nu_G$ requires weighting each pixel by $1/\sigma_x^2$ before the cross-correlation, which breaks the convolution structure and precludes efficient FFT-based evaluation. Following \citet{Zackay2017}, we adopt the locally stationary approximation: if $\sigma_x$ varies on spatial scales much larger than the kernel support, then $\sigma_x \approx \sigma_\mu$ within a region of radius $\sim\sigma_G$ centered on $\mu$. Under this approximation, the significance reduces to
\begin{equation}
    \nu_G(\mu) = 
    \frac{(M * \bar{G})_\mu}
    {\sigma_\mu \sqrt{A_G}},
    \label{eq:conv_sigma_isolated}
\end{equation}
where $\bar{G}_x \equiv G_{-x}$ is the reflected kernel and $A_G \equiv \sum_x G_x^2$ is the effective kernel area. In this form, the cross-correlation $(M * \bar{G})$ can be evaluated via FFT. The sensitivity loss from using a fixed Gaussian rather than the true PSF is negligible for the bright, high signal-to-noise (S/N) sources required for PSF modeling. The detection loss is only a few per
cent for a moderate ($\lesssim20\%$) width mismatch \citep[Appendix~D of][]{Zackay2017}. Local maxima of $\nu_G$ above a preliminary threshold of $N_{\sigma,G} = 5$ are extracted as initial candidate detections, shown as red circles in the top panel of Figure~\ref{fig:psf_isolated}.

Initial candidates are subsequently filtered by three criteria applied in sequence. First, an isolation criterion requires that each candidate lies at least $10$ arcsec from its nearest neighbor. Nearest-neighbor distances are computed via a $k$-d tree \citep{Bentley1975, Friedman1977}, which is efficient enough for the typical number of candidates ($N \lesssim 10^3$). Second, a stricter significance threshold $N_{\sigma,\mathrm{iso}} = 10$ is applied to $\nu_G$ to retain only high-confidence detections. Third, a saturation criterion rejects candidates whose peak pixel value exceeds 80 \% of the detector full-well capacity, providing a conservative margin against non-linear response. Sources close to the image edges are additionally excluded, ensuring that the extraction region of every retained candidate lies entirely within the detector. Candidates surviving all filters are marked in blue in Figure~\ref{fig:psf_isolated}, and represent the pool of bright, well-isolated point sources from which the PSF model is constructed.

\subsection{Reference PSF construction}
\label{sec:psf/reference}

From the isolated star candidates, a subset is selected to build a position-independent reference PSF $P_0$. Candidates are first ranked by a composite metric combining their signal-to-noise ratio $\nu_G$ with their nearest-neighbor distance, prioritizing bright, well-isolated sources. For each candidate, a square cutout of $10^{\prime\prime}$ on a side is extracted around its centroid. A cutout is retained only if the brightest pixel falls within $1^{\prime\prime}$ of the cutout center, and is then normalized by its total flux.

A quality filter is applied to identify stamps that are poorly sampled or anomalously compact, whose brightest pixel exceeds 30\% of the total stamp flux. Such stamps are dominated by a single bright pixel, which would bias the reference PSF towards sub-pixel artifacts rather than capturing the true profile. From the surviving stamps, only those lying within the central region are retained, where the PSF is typically most symmetric and least affected by field-dependent aberrations. The 21 highest-S/N central stamps are stacked with iterative sigma-clipping, and the resulting reference PSF is normalized to unit sum, $\sum_x (P_0)_x = 1$. The reference PSF is shown in the bottom-left panel of Figure~\ref{fig:psf_isolated}.

\subsection{Spatial variations via PCA}
\label{sec:psf/pca}

The reference PSF $P_0$ constructed in Section~\ref{sec:psf/reference} can be used directly as the matched-filter kernel for source detection. Substituting $P_0$ for $G$ in Equation~(\ref{eq:conv_sigma_isolated}) yields the analogous detection statistic,
\begin{equation}
    \nu_{P_0}(\mu) = \frac{(M * \bar{P}_0)_\mu}{\sigma_\mu \sqrt{A_{P_0}}},
    \label{eq:conv_sigma_ref}
\end{equation}
where $\bar{P}_0(x) \equiv P_0(-x)$ is the reflected reference PSF and $A_{P_0} \equiv \sum_x P_0^2(x)$ is its effective area. Equation~(\ref{eq:conv_sigma_ref}) is the optimal matched filter when the PSF is uniform across the image.

In practice, the PSF varies systematically across the field of view: optical aberrations, detector tilt, and atmospheric seeing gradients all introduce position-dependent deformations. Treating $P_0$ as the kernel everywhere therefore induces a position-dependent mismatch between the assumed and the true PSF, which degrades both the detection sensitivity and the photometric accuracy. Equation~(\ref{eq:conv_sigma_ref}) cannot be generalized by simply replacing $P_0$ by a position-dependent kernel $P(x; \mu)$, because the resulting operation is no longer a convolution: a kernel that varies with $\mu$ does not admit the FFT-based evaluation on which efficient detection relies.

To resolve this tension, we model the spatially varying PSF as a sum of fixed templates with position-dependent amplitudes,
\begin{equation}
    P(x; \mu) = P_0(x) + \sum_{k=1}^{K} c_k(\mu) e_k(x),
    \label{eq:psf_pca}
\end{equation}
where the basis functions $\{e_k\}_{k=1}^{K}$ (hereafter PSF deviation eigenmodes) are fixed templates determined empirically from the data, and the scalar coefficient maps $c_k(\mu)$ encode all of the spatial variation. The matched-filter result, which for a position-dependent kernel is no longer a single convolution, then becomes
\begin{equation}
    (M \circledast  \bar{P})_\mu = (M * \bar{P}_0)_\mu + \sum_k c_k(\mu) (M * \bar{e}_k)_\mu,
    \label{eq:matched_filter_pca}
\end{equation}
which can be evaluated as a sum of $K + 1$ genuine convolutions (with the fixed kernels $P_0$ and $e_k$) multiplied by the precomputed coefficient maps $c_k(\mu)$. We introduce the distinct symbol $\circledast$ to emphasize that, unlike the convolution $*$, this operation is not formally a convolution when the kernel $P$ varies with position.

\paragraph{Construction of the eigenmode basis}
For each isolated star $i$ found in the image, we compute its deviation from the reference PSF,
\begin{equation}
    d_i(x) \equiv \hat{s}_i(x) - P_0(x).
    \label{eq:psf_deviation}
\end{equation}
where $\hat{s}_i(x)$ is the unit-sum--normalized stamp of the $i$-th detected isolated star. Because both $\hat{s}_i$ and $P_0$ are normalized to unit sum (Sections~\ref{sec:psf/reference} and~\ref{sec:psf}), the deviations satisfy
\begin{equation}
    \sum_x d_i(x) = \sum_x \hat{s}_i(x) - \sum_x P_0(x) = 0.
    \label{eq:psf_deviation_zerosum}
\end{equation}
This zero-sum property is preserved under any linear combination of the $\{d_i\}$, and is the foundation of the flux-conservation argument given below.

\begin{figure}[t!]
\centering
\includegraphics[width=\linewidth]{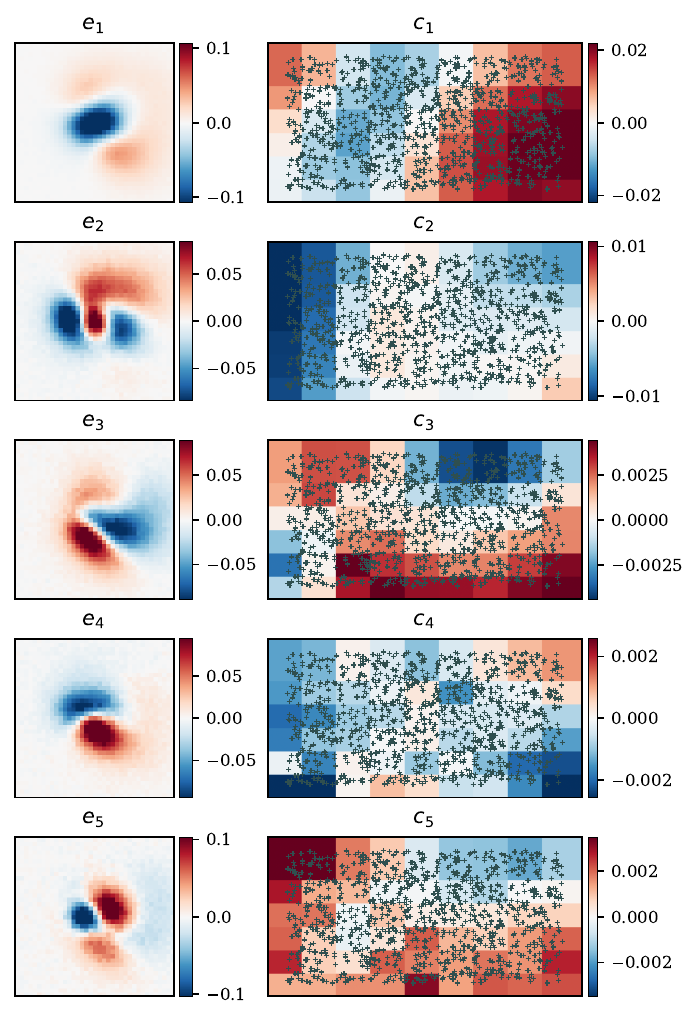}
\caption{PSF deviation eigenmodes and their spatial coefficient maps. \textit{Left column:} the first five eigenmodes $e_1, \ldots, e_5$ obtained from the weighted PCA of the noise-normalized PSF deviations $\tilde{d}_i$, ordered by decreasing variance explained. The color scale is symmetric around zero, with red and blue indicating positive and negative contributions to the local PSF relative to the reference $P_0$. \textit{Right column:} the corresponding interpolated coefficient maps $c_1(\mu), \ldots, c_5(\mu)$ across the full field of view, with the positions of the PSF modeling stars overplotted as gray crosses. The coherent large-scale structure in the maps indicates that the eigenmodes capture genuine spatial variations of the PSF rather than stochastic noise.
\label{fig:psf_pca}}
\end{figure}

A direct application of Principal Component Analysis (PCA) to the raw deviations $\{d_i\}$ would be dominated by the stars with the largest-amplitude deviations. Because the unit-sum normalization inflates the photon noise of faint stars, these are precisely the low-signal-to-noise stars whose deviations are dominated by noise, which would then contaminate the leading components. To mitigate this, we apply a weighted-PCA approximation: each deviation is first normalized by its own standard deviation,
\begin{equation}
    \tilde{d}_i(x) = \frac{d_i(x)}{\sigma_i},\quad \sigma_i \equiv \mathrm{std}_x\!\left[d_i(x)\right],
    \label{eq:psf_dev_norm}
\end{equation}
producing a set of unit-variance deviations in which each star contributes with comparable weight. PCA is then performed on the normalized set. The goal is to find a small set of fixed templates that, combined linearly, reproduce the bulk of the star-to-star PSF variation. Concretely, the eigenmodes $\{e_k\}_{k=1}^{K}$ are the $K$ leading principal components of the centered deviations $\{\tilde{d}_i - \langle\tilde{d}\rangle\}$: a set of mutually orthonormal patterns, ordered so that each successively captures the largest possible fraction of the remaining sample variance. The eigenmodes are, then, ordered by decreasing explained variance and satisfy
\begin{equation}
    \sum_x e_k(x) e_l(x) = \delta_{kl}.
    \label{eq:psf_orthonormality}
\end{equation}
\textsc{GPUPhot} retains the first $K = 5$ eigenmodes by default. Examples for a representative image are shown in the left column of Figure~\ref{fig:psf_pca}.

\paragraph{Projection coefficients and spatial interpolation}

Each star's deviation is projected onto the eigenmode basis. In the normalized space, the coefficient of $\tilde{d}_i$ along $e_k$ is
\begin{equation}
    \tilde{c}_{k,i} = \sum_x \tilde{d}_i(x) e_k(x).
    \label{eq:psf_projection}
\end{equation}
To recover coefficients in the physical (unnormalized) deviation space, we rescale by $\sigma_i$,
\begin{equation}
    c_{k,i} = \sigma_i \tilde{c}_{k,i},
    \label{eq:psf_coeff_rescale}
\end{equation}
so that the original deviation is recovered as $d_i(x) \approx \sum_k c_{k,i} e_k(x)$ up to the truncation residual associated with $K < N$.

The coefficients $\{c_{k,i}\}$ are known only at the discrete positions $\{\mu_i\}$ of the PSF modeling stars. To evaluate the PSF at any position, we interpolate each set of coefficients into a continuous spatial map $c_k(\mu)$. The image is divided into square tiles of side length $200$ arcsec by default, and within each tile the contributing stars are combined with a sigma-clipped mean to produce a single tile-averaged coefficient. Tiles without contributing stars are filled iteratively by FFT-based interpolation from their neighbors. The resulting maps $c_k(\mu)$ are piecewise-constant on the tile grid; the assumption that the PSF varies smoothly on scales larger than the tile size is justified by the smooth nature of the underlying optical and atmospheric variations. The coefficient maps for the same image are shown in the right column of Figure~\ref{fig:psf_pca}.

\paragraph{Flux conservation of the reconstructed PSF}
By construction, the deviations $d_i$ satisfy $\sum_x d_i(x) = 0$ (Equation~\ref{eq:psf_deviation_zerosum}), and so do the normalized deviations $\tilde{d}_i$ and their per-pixel mean $\langle\tilde{d}\rangle$. The eigenmodes $e_k$ are principal components of the centered deviations $\{\tilde{d}_i - \langle\tilde{d}\rangle\}$, hence linear combinations of zero-sum patterns, and therefore inherit the zero-sum property,
\begin{equation}
    \sum_x e_k(x) = 0,\quad k = 1, \dots, K.
    \label{eq:psf_eigenmode_zerosum}
\end{equation}
Substituting Equation~(\ref{eq:psf_eigenmode_zerosum}) into the reconstructed PSF model (Equation~\ref{eq:psf_pca}),
\begin{equation}
    \sum_x P(x;\mu) = \sum_x P_0(x) + \sum_k c_k(\mu) 
    \underbrace{\sum_x e_k(x)}_{= 0} = 1,
    \label{eq:psf_flux_conservation}
\end{equation}
so that the unit-sum normalization is preserved at every position $\mu$. This is essential for the matched-filter detection statistic, which assumes a flux-normalized PSF, and for the consistency of the photometric calibration.

\begin{figure}[t!]
\centering
\includegraphics[width=\linewidth]{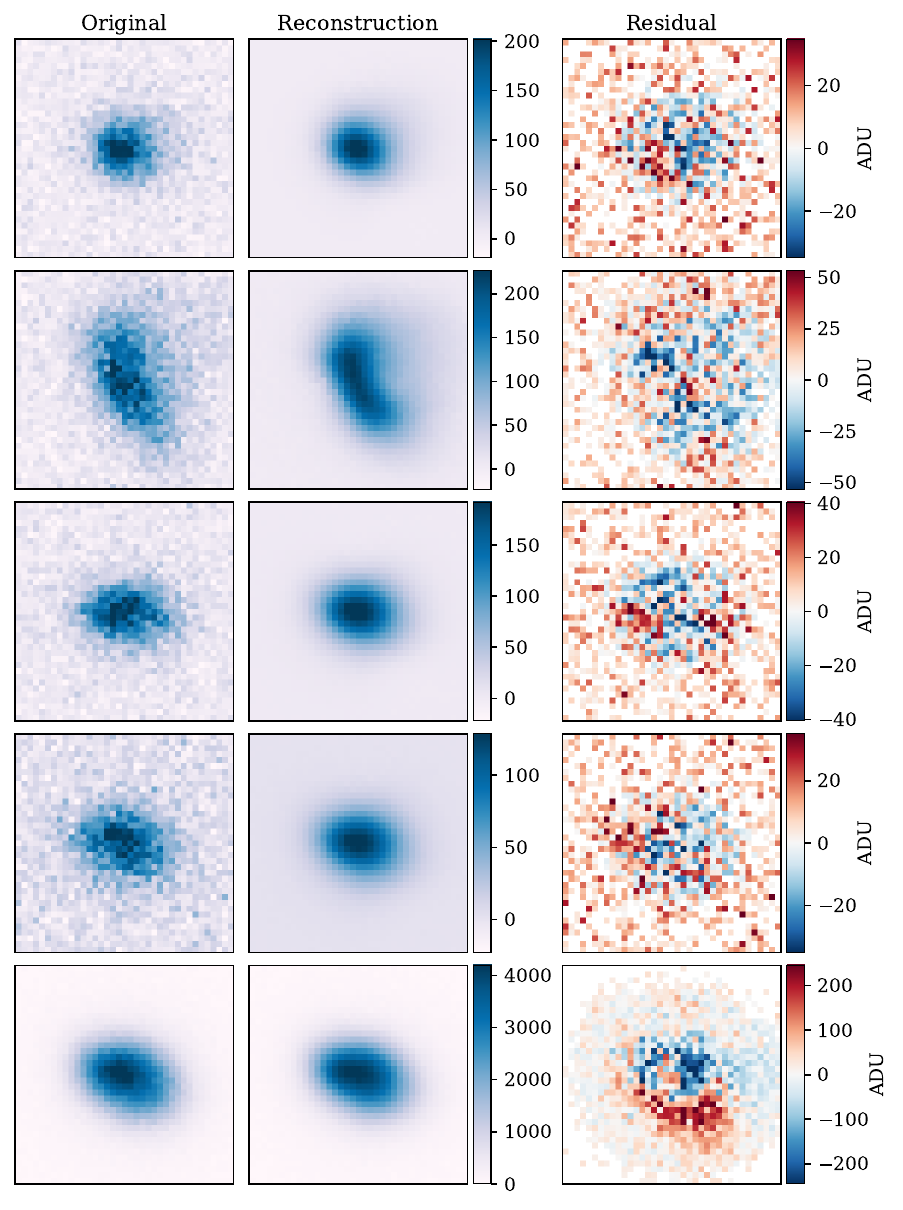}
\caption{Validation of the PCA-based PSF model. Each row corresponds to a star randomly drawn from the PSF modeling set. \textit{Left column:} the observed, background-subtracted stamp $s_i$ at the star's position, in detector units. \textit{center column:} the reconstructed PSF evaluated at the same position and rescaled by the star's total flux,
$F_i P(x; \mu_i)$ with $P(x; \mu_i) = P_0(x) + \sum_k c_k(\mu_i) e_k(x)$, where the reference PSF $P_0$ provides the bulk of the profile and the eigenmodes $e_k$, scaled by their local coefficients $c_k(\mu_i)$, add the position-dependent correction that captures the PSF variation across the field. \textit{Right column:} the residual $ [s_i - F_i P(x; \mu_i)]$, shown over the source region on a symmetric color scale.
\label{fig:psf_residuals}}
\end{figure}

\paragraph{Validation}
Figure~\ref{fig:psf_residuals} shows representative reconstructions for stars distributed across the field: the observed stamp $s_i$, the flux-scaled model PSF $F_i P(x; \mu_i)$ and the residual $s_i - F_i P(x; \mu_i)$. For all but the brightest star the residuals exhibit no systematic structure beyond pixel-level photon noise. This shows a faint dipolar residual, a positive and a negative lobe on opposite sides of the center, which is the first-order signature of a small, sub-pixel offset between the model and the stellar centroid, rather than a deficiency of the eigenmode basis. Such an offset is present at some level for every star but is masked by photon noise except at the highest signal-to-noise, where the residual is clean enough for it to become visible; it reflects the limiting positional accuracy rather than the PSF shape, is absorbed in the per-source position fit, and does not affect the shape fidelity tested here. The eigenmode basis with $K = 5$ thus captures the spatial PSF variation with sufficient fidelity for accurate detection across the full field of view.

\begin{figure*}[ht!]
\centering
\includegraphics[width=\linewidth]{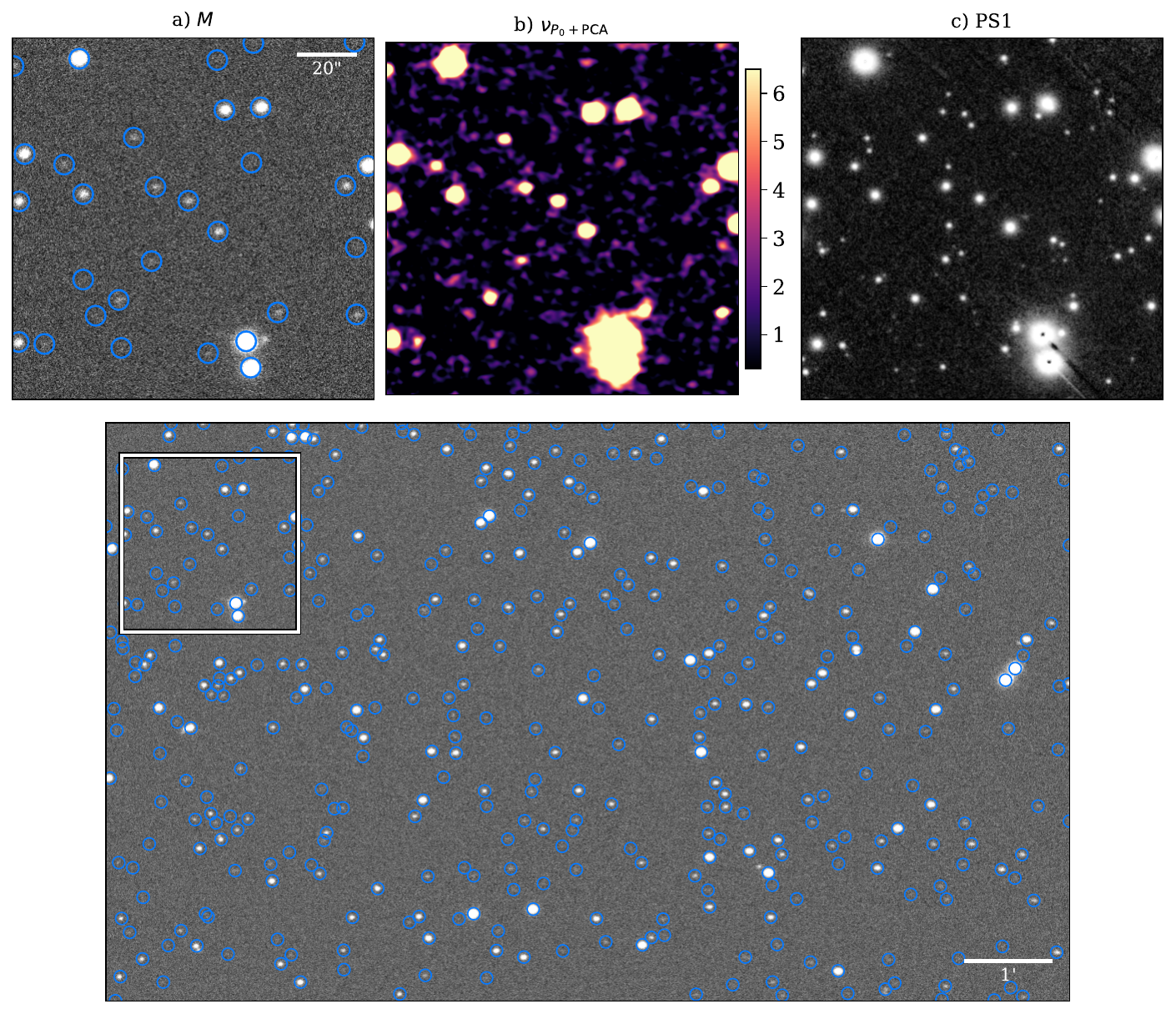}
\caption{Source detection on a representative field. \textit{Top row:} a small cutout extracted from the full detection region, shown from three perspectives. \textbf{a)} background-subtracted image $M$, with the final sub-pixel-refined source positions overlaid as blue circles; \textbf{b)} matched-filter significance map $\nu_{P_0 + \mathrm{PCA}}$ computed with the full spatially varying PSF model; \textbf{c)} corresponding Pan-STARRS DR1 image \citep{Chambers2016}, included as an external reference to verify that the detected sources correspond to genuine astronomical objects. \textit{Bottom row:} \textbf{(d)} the complete detection cutout with all sources marked as blue circles; the white rectangle indicates the location of the small cutout shown in panels a)--c). The horizontal bars indicate angular scales of $20^{\prime\prime}$ in the top row and $1^\prime$ in the bottom row. The agreement between the \textsc{GPUPhot} detections and the deeper Pan-STARRS imagery confirms that the algorithm recovers the faint sources visible in the reference survey, while introducing no spurious detections beyond those expected from photon noise.
\label{fig:detection_pipeline}}
\end{figure*}

\section{Source detection}
\label{sec:detection}

With the spatially varying PSF model in place, source detection in \textsc{GPUPhot} reduces to applying the matched filter of Equation~(\ref{eq:matched_filter_pca}) and identifying local maxima of the resulting significance map. The procedure is the natural generalization of Section~\ref{sec:psf/isolated} from a fixed Gaussian kernel to the empirical PSF, retaining the same maximum-likelihood foundation under the locally stationary noise approximation.

The detection statistic at each pixel is
\begin{equation}
    \nu(\mu) = \frac{(M \circledast \bar{P})_\mu}{\sigma_\mu \sqrt{A(\mu)}},
    \label{eq:detection_S/N_pca}
\end{equation}
where $(M \circledast \bar{P})_\mu$ is the spatially varying matched filter of Equation~(\ref{eq:matched_filter_pca}), and the position-dependent effective kernel area is
\begin{equation}
    A(\mu) = \sum_x P^2(x;\mu) = A_{P_0}
    + 2\sum_{k=1}^{K} c_k(\mu)\,\pi_k
    + \sum_{k=1}^{K} c_k^2(\mu),
    \label{eq:eff_area_pca}
\end{equation}
where $\pi_k \equiv \sum_x P_0(x)e_k(x)$ are the projections of the reference PSF onto the eigenmode basis. The orthonormality of the eigenmodes (Equation~\ref{eq:psf_orthonormality}) cancels the cross terms $\sum_x e_k(x)e_l(x)$ for $k \neq l$ and sets $\sum_x e_k^2(x) = 1$. It does not cancel the projections $\pi_k$: the basis is orthonormal within itself and is not orthogonalized against $P_0$. The $\pi_k$ are $K$ scalars evaluated once per frame,
so Equation~(\ref{eq:eff_area_pca}) adds no measurable cost. Both the numerator and the effective area are therefore obtained entirely from FFT-based convolutions: the matched filter as a sum of $K+1$ fixed-kernel convolutions multiplied by the precomputed coefficient maps, and the effective area from the same maps and the precomputed scalars $\pi_k$.

Local maxima of $\nu$ above a configurable significance threshold $N_\sigma$ (default $N_\sigma = 4$) are identified as source candidates. Two candidates are required to be separated by at least $2 \times \mathrm{FWHM}$ pixels to avoid duplicate detections of the same source split by noise fluctuations. The integer-pixel peak positions are subsequently refined to sub-pixel accuracy by computing the first-moment centroid of $\nu$ within a square window of side $1 \times \mathrm{FWHM}$ around each maximum,
\begin{equation}
\hat{\mu} = \mu + \frac{\sum_{x' \in W} \nu(x') (x' - \mu)}{\sum_{x' \in W} \nu(x')},
    \label{eq:centroid}
\end{equation}
where $x'$, $\mu$ and $\hat{\mu}$ are two-dimensional position vectors and $W$ is the centroid window.

Figure~\ref{fig:detection_pipeline} illustrates the detection on a representative field. The top row shows a small cutout extracted from the full field, with the same region rendered as: the background-subtracted image $M$ with the final source detections overlaid (panel a), the matched-filter significance map $\nu_{P_0 + \mathrm{PCA}}$ (panel b), and the corresponding Pan-STARRS DR1 image \citep{Chambers2016} as an external reference (panel c). The bottom panel (d) shows the full detection cutout with all sources marked. The Pan-STARRS comparison confirms that the algorithm recovers the faint sources visible in the deeper reference survey while introducing no spurious detections beyond what is expected from photon noise.

\section{Photometry}
\label{sec:photometry}

Once all detected sources have been assigned a sub-pixel position by the centroid refinement of Equation~(\ref{eq:centroid}), \textsc{GPUPhot} measures their flux by optimized circular aperture photometry. The signal at each detected source is the sum of the pixel values inside a circular aperture of radius $r$ centered on the refined position $\hat{\mu}$, plus a small fraction of the flux that falls outside the aperture and which has to be recovered. The procedure can be cleanly decomposed into four parts: the aperture flux measurement itself, the aperture correction that recovers the missing flux, the noise model that propagates the photometric uncertainty and the brightness-dependent rule that selects the optimal aperture radius for each source.

\subsection{Aperture flux measurement}
\label{sec:photometry/flux}

For a given aperture radius $r$, the measured flux of the source at position $\hat{\mu}$ is the inner product of the background-subtracted image $M$ with a circular top-hat kernel $K_r$ of unit weight inside radius $r$ and zero outside,
\begin{equation}
    F(r;\hat{\mu}) \equiv \sum_x M_x K_r(x - \hat{\mu}), \quad
    K_r(x) =
    \begin{cases}
        1, & |x| \leq r \\
        0, & |x| > r
    \end{cases}
    \label{eq:aperture_flux_definition}
\end{equation}
Apertures are pixel-discrete: a pixel is included in $K_r$ if and only if its integer coordinates satisfy $|x|\le r$, with no sub-pixel weighting at the aperture boundary. This is a deliberate choice that preserves the matched-filter structure of the rest of the pipeline and allows $F(r;\hat{\mu})$ to be evaluated efficiently for all sources simultaneously via FFT-based cross-correlation,
\begin{equation}
    F(r;\hat{\mu}) = (M * \bar{K}_r)_{\hat{\mu}},
    \label{eq:aperture_flux_fft}
\end{equation}
where $\bar{K}_r(x) \equiv K_r(-x) = K_r(x)$ by the symmetry of the top-hat. The same FFT pathway used in Sections~\ref{sec:psf/isolated} and~\ref{sec:detection} therefore applies, with $K_r$ playing the role of the matched-filter kernel. %Because the cross-correlation is evaluated on the pixel grid, the aperture is in practice centered on the pixel nearest to $\hat{\mu}$ rather than on $\hat{\mu}$ itself. The residual offset is bounded by half a pixel and, together with the discrete aperture boundary, introduces a deficit that depends on the sub-pixel phase of the source. Its magnitude is bounded by the discreteness of the aperture and is absorbed, in the mean, into the photometric zero point.

In practice, the algorithm does not evaluate Equation~(\ref{eq:aperture_flux_fft}) at a single radius but at a discrete set of integer radii, in pixels, spanning a range tailored to the local PSF width,
\begin{equation}
    \lceil 0.75 \times \mathrm{FWHM} \rceil \leq r \leq \lceil 5 \times \mathrm{FWHM} \rceil, \quad r \in \mathbb{N}
\label{eq:aperture_radii_set}
\end{equation}
producing for each source a growth curve $F(r;\hat{\mu})$ sampled at $N_r$ radii. The lower bound is set such that the smallest aperture comfortably contains the FWHM core of the PSF, while the upper bound extends well into the wings, beyond the radius at which the aperture correction stabilizes.
The area of each aperture is computed exactly from the same top-hat,
\begin{equation}
    A(r) = \sum_x K_r(x),
    \label{eq:aperture_area}
\end{equation}
and enters the noise model as the number of pixels contributing readout noise and dark current. Because $K_r$ is rotationally symmetric and shared across all sources, both $F(r;\hat{\mu})$ and $A(r)$ depend only on $r$ and the source position, not on the local PSF shape; all PSF-dependent corrections are deferred to the aperture-correction step that follows.

\subsection{Aperture correction}
\label{sec:photometry/apercorr}

\begin{figure}[t!]
\centering
\includegraphics[width=\linewidth]{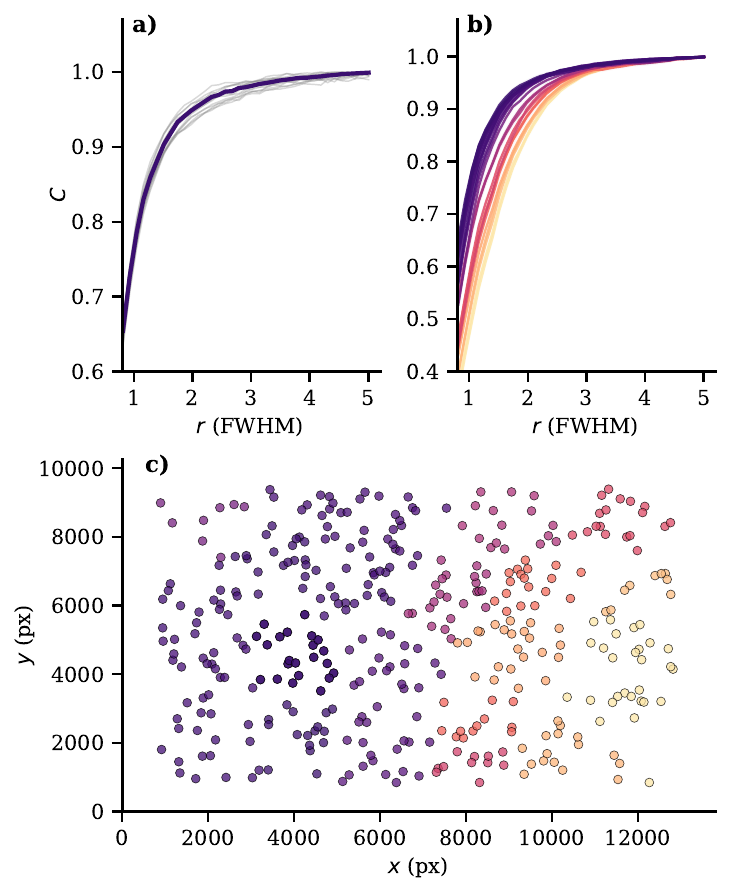}
\caption{Empirical aperture correction across the field of view. \textbf{a)} Within-cluster growth curves for a single representative spatial cluster: individual normalized growth curves $\hat{C}_i(r)$ of the $n_c$ retained calibrators (gray), overlaid by the cluster-mean correction $C_c(r)$ (purple). \textbf{b)} Cluster-mean corrections $C_c(r)$ for all $N_c$ clusters in the field, overlaid on a common axis and colored by their value at one FWHM. \textbf{c)} Spatial distribution of the calibrators in the focal plane, each colored by the $C_c(\mathrm{FWHM})$ value of its parent cluster.
\label{fig:aperture_corrections}}
\end{figure}

The aperture flux $F(r;\hat{\mu})$ defined in Equation~(\ref{eq:aperture_flux_definition}) systematically underestimates the total source flux because a finite fraction of the PSF lies outside any finite aperture. We define the aperture correction $C(r;\mu)$ at position $\mu$ as the fraction of the total flux of a unit-normalized PSF that is enclosed within radius $r$,
\begin{equation}
    C(r;\mu) \equiv \sum_{|x| \leq r} P(x;\mu),
    \label{eq:aperture_correction_definition}
\end{equation}
so that the calibrated signal of a source at $\hat{\mu}$ with measured aperture flux $F(r;\hat{\mu})$ is
\begin{equation}
    S(r;\hat{\mu}) = \frac{F(r;\hat{\mu})}{C(r;\hat{\mu})}.
    \label{eq:calibrated_signal}
\end{equation}
The quantity $S$ is an estimator of the true point-source flux $\alpha$ introduced in Equation~(\ref{eq:point_source_model_general}): by construction, $C(r;\mu) \to 1$ as $r$ grows beyond the PSF support, so $S$ recovers $\alpha$ in the limit of large apertures.

Direct evaluation of Equation~(\ref{eq:aperture_correction_definition}) from the PCA-based PSF model of Section~\ref{sec:psf/pca} is in principle possible, but aperture photometry imposes far tighter accuracy requirements on $C(r;\mu)$ than source detection does. The matched-filter detection statistic of Section~\ref{sec:detection} depends on the global convolution $(M * \bar{P})$ and tolerates percent-level errors in the radial profile of $P$ without noticeable degradation, whereas the calibrated signal of Equation~(\ref{eq:calibrated_signal}) and the optimal-aperture selection of Section~\ref{sec:photometry/optradius} both depend directly on the radial integral and propagate any such error into a percent-level systematic in the final flux. We therefore adopt an empirical approach that bypasses Equation~(\ref{eq:aperture_correction_definition}) altogether and estimates $C(r;\mu)$ directly from observed growth curves of isolated stars in the field, which provides the required accuracy by construction.

The procedure has three stages: selection of high-quality aperture calibrators (high-S/N isolated stars), segmentation of the field into spatial clusters, and estimation of the cluster-mean correction together with its uncertainty. The result, illustrated in Figure~\ref{fig:aperture_corrections} for a representative field, is a set of $N_c$ piecewise-constant correction curves $C_c(r)$ that capture the smooth spatial PSF variation across the focal plane.

The starting point are the isolated stars detected in Section~\ref{sec:psf/isolated}, of which a subset is retained as aperture-correction calibrators by requiring a convolutional signal-to-noise (Equation~\ref{eq:detection_S/N_pca}) exceeding $\nu = 300$, automatically relaxed to $100$ if the resulting sample contains fewer than ten stars. The high threshold ensures that the growth curve of each calibrator is photon-noise limited well into the PSF wings, where the small flux increments per pixel are otherwise dominated by background fluctuations. For each calibrator $i$ at position $\mu_i$, the growth curve $F_i(r) \equiv F(r;\mu_i)$ is measured on a discrete grid extending to $r_\mathrm{norm} = 6 \times \mathrm{FWHM}$, beyond the photometric range of Equation~(\ref{eq:aperture_radii_set}), and normalized by its own maximum value over that extended range,
\begin{equation}
    \hat{C}_i(r) \equiv \frac{F_i(r)}{\max_{r'} F_i(r')}.
    \label{eq:growth_curve_normalized}
\end{equation}
By construction each $\hat{C}_i$ reaches unity at the radius where its own growth curve peaks, so the within-cluster dispersion collapses in the neighbourhood of $r_\mathrm{norm}$. Normalizing over the photometric range alone would place this artificial zero at $r = r_\mathrm{max}$, where it would erase the systematic term from the noise budget and bias the optimal-radius selection of the following section. Extending the normalization to $r_\mathrm{norm} = 6\,\mathrm{FWHM}$ pushes it safely outside the photometric range, leaving $\sigma_{C_c}(r)$ finite and informative at every radius actually used for photometry.

The calibrators are then grouped into spatial clusters that divide the field of view. The target number of clusters $N_c$ is set to
\begin{equation}
    N_c = \left\lfloor N_\mathrm{cal} \cdot \frac{T^2}{H W}\right\rfloor,
    \label{eq:n_clusters}
\end{equation}
where $T$ is the same tile size used for the PCA coefficient maps, $N_\mathrm{cal}$ is the total number of retained calibrators, and $H \times W$ is the image size in pixels. This choice adapts naturally to image quality: a sparsely populated or shallow image yields fewer calibrators and therefore fewer clusters, each averaging over a larger area and a more stable mean correction. A floor of $n_\mathrm{min} = 5$ calibrators per cluster is enforced so that the within-cluster statistics remain meaningful. Calibrators are assigned to clusters by a $k$-means partition of their pixel coordinates, and within each cluster $c$, outliers more than $2.5\,\mathrm{MAD}$ from the median at each radius are iteratively rejected. Panel c) of Figure~\ref{fig:aperture_corrections} shows the resulting spatial partition for a representative field: the calibrators are colored by the index of their parent cluster.
The cluster-mean correction and its uncertainty are then
\begin{equation}
\begin{aligned}
    C_c(r) =& \left\langle \hat{C}_i(r) \right\rangle_{i \in c} \\
    \sigma_{C_c}(r) =& \mathrm{std}_{i \in c}\!\left[\hat{C}_i(r)\right] \big/ \sqrt{n_c},
\end{aligned}
\label{eq:cluster_correction}
\end{equation}
where $n_c$ is the number of retained calibrators in the cluster. Two effects contribute to the scatter of the $\hat{C}_i(r)$ within a cluster: the photon noise of the individual growth curves, which is independent between calibrators, and any residual PSF variation across the cluster area. The second is small by construction, because the number of clusters of Equation~(\ref{eq:n_clusters}) is set so that each cluster spans approximately one PCA tile, i.e.\ the scale over which the PSF is already assumed to be constant in Section~\ref{sec:psf/pca}; the spread between clusters shown in panel~b) of
Figure~\ref{fig:aperture_corrections} is the fingerprint of the variation that the partition does capture. The dispersion is therefore dominated by the independent photon-noise term and averages down with $n_c$, which is what Equation~(\ref{eq:cluster_correction}) assumes.

Panel a) of Figure~\ref{fig:aperture_corrections} illustrates the within-cluster statistics for a single representative cluster: the individual normalized growth curves $\hat{C}_i(r)$ of the $n_c$ retained calibrators are shown in gray, and the cluster-mean correction $C_c(r)$ is overlaid in black. The mean curve approaches unity well within the photometric range, confirming that the upper bound $r_\mathrm{max} = 5\,\mathrm{FWHM}$ adopted in Equation~(\ref{eq:aperture_radii_set}) encompasses essentially the full source flux. Panel b) overlays the cluster-mean corrections of all $N_c$ clusters in the field on a common axis, each curve colored by the index of its cluster as in panel c). Because the top-hat kernel $K_r$ introduced in Equation~(\ref{eq:aperture_flux_definition}) is identical for all sources, the entire spatial variation of the PSF is absorbed into a position-dependent correction $C(r;\mu)$. The non-trivial spread between clusters in panel b) is the direct fingerprint of that variation: as anticipated from the PCA eigenmodes of Section~\ref{sec:psf/pca}, the PSF core varies more strongly across the field than its wings, so the curves diverge at small radii and converge at large ones.

In the third stage, every source at position $\hat{\mu}$ is assigned the correction of its nearest cluster centroid.

\subsection{Noise contributions}
\label{sec:photometry/noise}

\begin{figure}[t!]
\centering
\revisiongraphic[width=\linewidth]{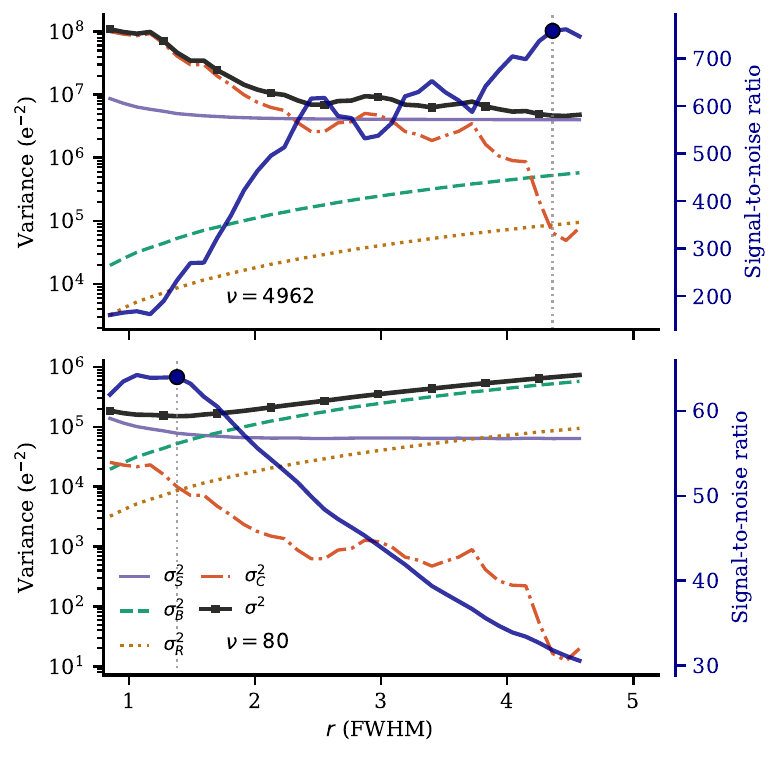}
\caption{Noise contributions as a function of aperture radius for two representative isolated stars in the same field. \textit{Top:} a bright star with detection significance $\nu \sim 5000$. \textit{Bottom:} a faint star with $\nu \sim 80$. Each panel shows the individual variance contributions to the calibrated signal in electrons: source Poisson ($\sigma_{S}^2$, solid), sky background ($\sigma_{B}^2$, dashed), readout ($\sigma_{R}^2$, dotted) and aperture-correction systematic ($\sigma_{C}^2$, dash-dotted); the total variance is overlaid in black with square symbols. The signal-to-noise ratio is shown on the secondary axis (blue, right) with the maximum marked as a filled circle.
\label{fig:noise_model}}
\end{figure}

The total noise on the aperture flux is the quadrature sum of four physically independent contributions: the source Poisson noise, the sky background Poisson noise, the readout noise and the dark current shot noise,
\begin{equation}
    \sigma_{\rm tot}^2(r) = \sigma_{S}^2(r) + \sigma_{B}^2(r) + \sigma_{R}^2(r) + \sigma_{D}^2(r),
    \label{eq:noise_total}
\end{equation}
all expressed in electrons. Each term is evaluated in a way appropriate to its spatial structure. Contributions that can be treated as spatially uniform across the image, such as the readout noise for a single-channel CCD, reduce to a single per-pixel variance multiplied by the area $A(r)$ of the aperture defined in Equation~(\ref{eq:aperture_area}). Contributions that vary with position, such as the sky background or the dark current noise, are instead obtained by convolving the corresponding per-pixel variance map with the same top-hat kernel $K_r$ used to measure the aperture flux $F(r;\hat{\mu})$ in Equation~(\ref{eq:aperture_flux_fft}), so that the variance of each term shares the same FFT-based machinery as the photometry itself.

The source noise has two physically distinct contributions, both arising from the propagation of Equation~(\ref{eq:calibrated_signal}): the Poisson noise of the source photons collected within the aperture, amplified by the multiplicative correction, and the systematic uncertainty introduced by $C(r;\hat{\mu})$ itself. The second contribution becomes essential when considering which is the optimal aperture for measuring the total flux. If $C(r;\mu)$ were taken without uncertainty, the calibrated signal $S = F/C$ would always favor the smallest possible apertures, since shrinking the aperture reduces all measurement terms while a perfect correction recovers the missing flux at no cost. Empirically this is not what is observed. For a bright source, dominated by source-photon noise, larger apertures that capture essentially all of the flux yield more precise photometry than small apertures recovered through a multiplicative correction, because the correction is never perfect: it inherits the PSF variability across the field and the limitations of pixel sampling. For a faint source, dominated by background noise, the trade-off reverses, since the cost of integrating more pixels grows faster than the gain from including more source signal. Figure~\ref{fig:noise_model} illustrates this asymmetry by dissecting the noise contributions of two representative isolated stars in the same field, spanning more than two orders of magnitude in detection significance: the S/N curves on the secondary axes peak at very different radii on the two panels, large for the bright star and small for the faint one.

Carrying out the propagation explicitly, with $\sigma_F^2 = F$ from Poisson statistics, gives the source variance as
\begin{equation}
    \sigma_{S}^2(r) = \frac{F(r;\hat{\mu})}{C^2(r;\hat{\mu})} + \frac{F^2(r;\hat{\mu})\,\sigma_{C}^2(r;\hat{\mu})}{C^4(r;\hat{\mu})},
    \label{eq:noise_src}
\end{equation}
where $\sigma_C(r;\hat{\mu}) \equiv \sigma_{C_{c^*}}(r)$ is the per-cluster dispersion of Equation~(\ref{eq:cluster_correction}) for the cluster $c^*$ assigned to the source.

The signal-to-noise ratio at aperture radius $r$ follows as
\begin{equation}
    \mathrm{S/N}(r;\hat{\mu}) = \frac{S(r;\hat{\mu})}{\sigma(r;\hat{\mu})}.
    \label{eq:S/N_aperture}
\end{equation}

\subsection{Optimal aperture radius}
\label{sec:photometry/optradius}

The two regimes identified in Section~\ref{sec:photometry/noise} make the dependence of the S/N on aperture radius non-trivial, with the maximum shifting from small to large $r$ as source brightness increases. For each source, the optimal aperture is therefore the radius that maximizes the signal-to-noise ratio of Equation~(\ref{eq:S/N_aperture}).
Evaluating this for every detected source would require computing all four noise terms of Equation~(\ref{eq:noise_total}) at every aperture radius for every source, a calculation that is straightforward for the isolated stars used to construct the noise model, but expensive when extended to all the detected sources.

Instead, we exploit a quantity already available for every detection: the detection significance $\nu(\hat{\mu})$ of Equation~(\ref{eq:detection_S/N_pca}), illustrated in panel b) of Figure~\ref{fig:detection_pipeline}. This statistic captures the ratio between the source signal extracted by the matched filter and the local background noise. It is not equal to the photometric S/N of Equation~(\ref{eq:S/N_aperture}) --- $\nu$ is evaluated under the null hypothesis, so it scales linearly with flux, whereas the photometric S/N scales as $\sqrt{F}$ in the source-dominated regime --- but it is a strictly monotonic function of source brightness, computed once during detection, and this is what the empirical calibration below requires. The power-law form of Equation~(\ref{eq:r_opt_law}) absorbs the difference between the two scalings.

\begin{figure}[t!]
\centering
\includegraphics[width=\linewidth]{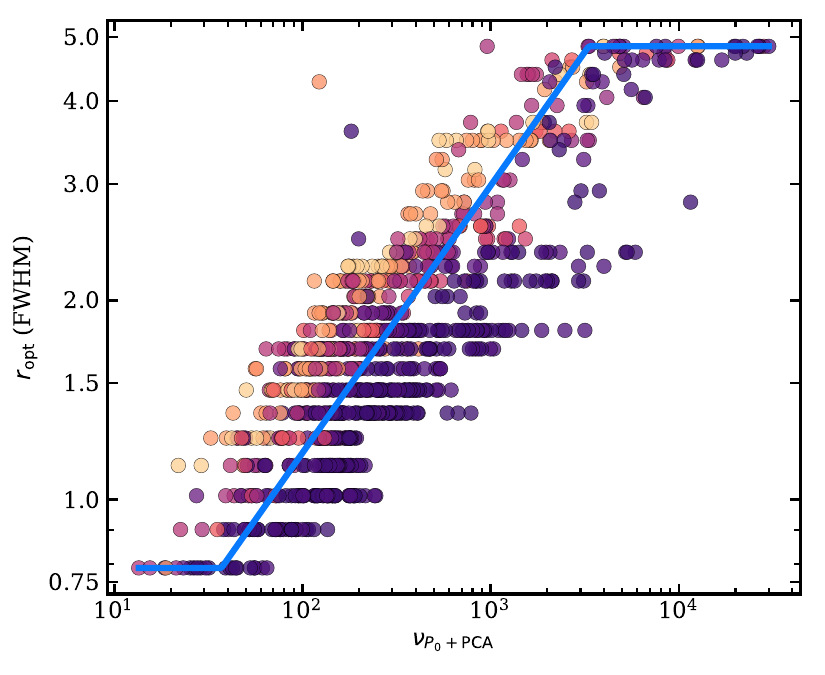}
\caption{Calibration of the optimal aperture radius. Optimal radius $r_\mathrm{opt}$ (in units of FWHM) as a function of the convolutional detection significance $\nu_{P_0+\mathrm{PCA}}$ for the population of isolated calibrators in a representative field. Each point is colored by the spatial cluster to which the source is assigned in the aperture-correction map of Section~\ref{sec:photometry/apercorr}, with the same color scheme as in Figure~\ref{fig:aperture_corrections}. The black solid line shows the inverted-binned power-law fit of Equation~(\ref{eq:r_opt_law}), with the saturated extremes at $r_\mathrm{min}$ and $r_\mathrm{max}$ excluded from the fit.
\label{fig:optimal_radius}}
\end{figure}

The optimal aperture radius can then be calibrated as a function of $\nu$. For each isolated star $i$, the photometric S/N curve is evaluated on the grid of Equation~(\ref{eq:aperture_radii_set}) and the radius at which it peaks is recorded as $r_{\mathrm{opt},i}$, as illustrated in Figure~\ref{fig:noise_model}. The resulting pairs $(\nu_i, r_{\mathrm{opt},i})$ trace a smooth relation across a wide brightness range, shown in Figure~\ref{fig:optimal_radius}. On the log--log plane, this relation is described by a power law,
\begin{equation}
    \log_{10} r_\mathrm{opt} = a\,\log_{10}\nu + b,
    \label{eq:r_opt_law}
\end{equation}
fitted to the isolated sources. Two practical aspects of the fit deserve mention. First, $r_\mathrm{opt}$ is a discrete-valued quantity by construction (an integer pixel radius from the grid of Equation~\ref{eq:aperture_radii_set}), so a direct ordinary-least-squares fit of $\log r_\mathrm{opt}$ on $\log\nu$ is biased: at fixed brightness, the scatter of $r_\mathrm{opt}$ is asymmetric because the radii cluster at integer values. We bypass this by binning the calibrators by integer $r_\mathrm{opt}$, computing the median $\log\nu$ within each bin. Second, calibrators whose $r_\mathrm{opt}$ saturates at either end of the photometric grid are excluded from the fit, since their $r_\mathrm{opt}$ is bounded by the grid rather than determined by the maximum S/N.

\begin{figure*}[t!]
\centering
\revisiongraphic[width=\linewidth]{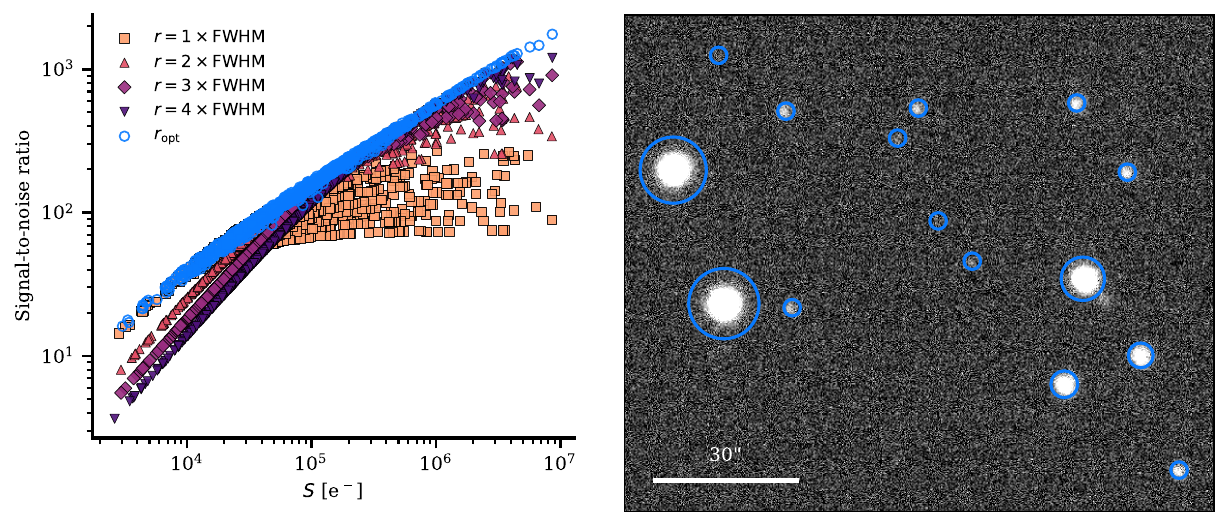}
\caption{Photometric performance of the adaptive aperture scheme. \textit{Left:} Photometric S/N as a function of total signal $S$ for the adaptive aperture (open blue circles) compared with four fixed-aperture alternatives at $r = 1, 2, 3, 4 \times$FWHM (red-to-purple gradient). The adaptive curve traces the upper envelope of the fixed alternatives across the full dynamic range. \textit{Right:} Representative cutout of the field with each detected source enclosed by a circle of radius $r_\mathrm{opt}(\hat{\mu})$. The aperture size adapts smoothly to source brightness, with the brightest stars receiving wide apertures that minimize the aperture-correction systematic and the faintest receiving compact apertures that limit background contamination.
\label{fig:photometric_performance}}
\end{figure*}

Once Equation~(\ref{eq:r_opt_law}) is calibrated, it is applied to every detected source by evaluating it at the convolutional significance $\nu(\hat{\mu})$ already available from detection, with the resulting radius rounded to the grid of Equation~(\ref{eq:aperture_radii_set}) and clipped to its boundaries. In this way, the optimal radius for every detection is determined from a single quantity already available at detection time, without recomputing the noise model on a source-by-source basis.

This procedure is validated in Figure~\ref{fig:photometric_performance}. The left-hand panel compares the photometric S/N as a function of the total signal $S$ for the adaptive aperture against four fixed-aperture alternatives. The adaptive curve traces the upper envelope of the fixed apertures across the full dynamic range: at the faint end it follows the small-aperture curves, where a wider aperture would incur an avoidable background penalty; at the bright end it converges with the larger ones, where the aperture-correction systematic is minimized by a near-unit $C$. No single fixed radius is competitive across the full brightness range, which is the basic motivation for the adaptive scheme. The right-hand panel illustrates the result on a representative cutout of the field: each detected source is enclosed by a circle of radius $r_\mathrm{opt}(\hat{\mu})$, with bright stars assigned wide apertures that capture the full PSF wings and faint stars assigned compact apertures that minimize background contamination. The photometric dataset produced by this scheme provides the per-source position, total flux $S(\hat{\mu})$ and uncertainty $\sigma(\hat{\mu})$.

\section{Results}
\label{sec:results}

\subsection{Synthetic injection--recovery validation}
\label{sec:results_injection}

Characterizing the performance of a source-extraction pipeline requires a ground truth against which its output can be compared. On real images the true positions and fluxes of the sources are unknown, so any assessment based on external catalogs is limited by the accuracy, depth and bandpass of the reference data, and cannot probe the faint end where the pipeline is most stressed. We therefore validate \textsc{GPUPhot} with a controlled injection--recovery experiment, in which synthetic point sources of known position and flux are inserted into images and recovered by running the full pipeline from scratch. This approach provides an exact truth catalog at every brightness, isolates the behavior of the estimators from calibration uncertainties, and allows the false-positive rate to be measured unambiguously.

We carry out this validation on four instruments chosen to span a wide range of plate scales and PSF sampling, so that the spatially varying PSF model is tested across regimes. The TTT-2m telescope, equipped with a CCD, provides a small, well-sampled field ($7.9\arcmin\times7.9\arcmin$ at $0.230\arcsec\,{\rm px}^{-1}$); the TTT-80cm telescope contributes a wider sCMOS field ($33\arcmin\times25\arcmin$ at $0.283\arcsec\,{\rm px}^{-1}$); the TST is a wide-field sCMOS imager ($2.4\degr\times1.8\degr$ at $0.597\arcsec\,{\rm px}^{-1}$) with a large field of view and strong PSF variation; and the ATLAS-Teide telescope provides an even wider, undersampled sCMOS field ($3.3\degr\times2.2\degr$ at $1.258\arcsec\,{\rm px}^{-1}$). Together they cover the full range from small, well-sampled fields to large, undersampled ones.

To avoid the confusion that real sources would introduce and to control the noise properties exactly, the synthetic images are built on a flat background rather than by injecting onto science frames. For each test image we inherit the measured sky level and the noise model of the corresponding real observation (Section~\ref{sec:background}), and construct a uniform background at the constant level. Point sources are then added using the spatially varying PSF model of Section~\ref{sec:psf}: for a source at position $(x,y)$ the reconstructed PSF $P(x;\mu)$ is evaluated at that location, shifted to the requested sub-pixel position with a third-order spline, and scaled to the injected flux. The sub-pixel barycenter of the shifted stamp is recorded as the true position, so that the astrometric residuals are measured against the actual center of light rather than the nominal coordinate.

\begin{figure*}
    \centering
    \begin{subfigure}{\textwidth}
        \centering
        \includegraphics[width=\textwidth]{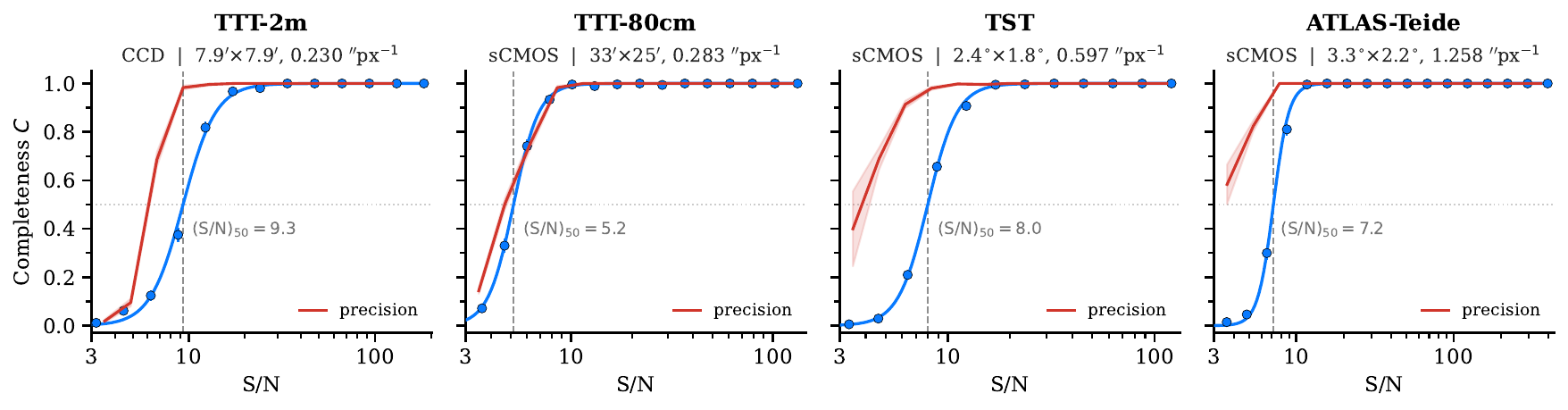}
        \caption{Completeness $C$ (blue) as a function of predicted S/N, with a logistic fit and the $(\mathrm{S/N})_{50}$ level marked; the precision (red; $\mathrm{TP}/(\mathrm{TP}+\mathrm{FP})$) quantifies the false-positive contamination measured on the same images.}
        \label{fig:res/completeness}
    \end{subfigure}

    \vspace{0.4em}
    \begin{subfigure}{\textwidth}
        \centering
        \revisiongraphic[width=\textwidth]{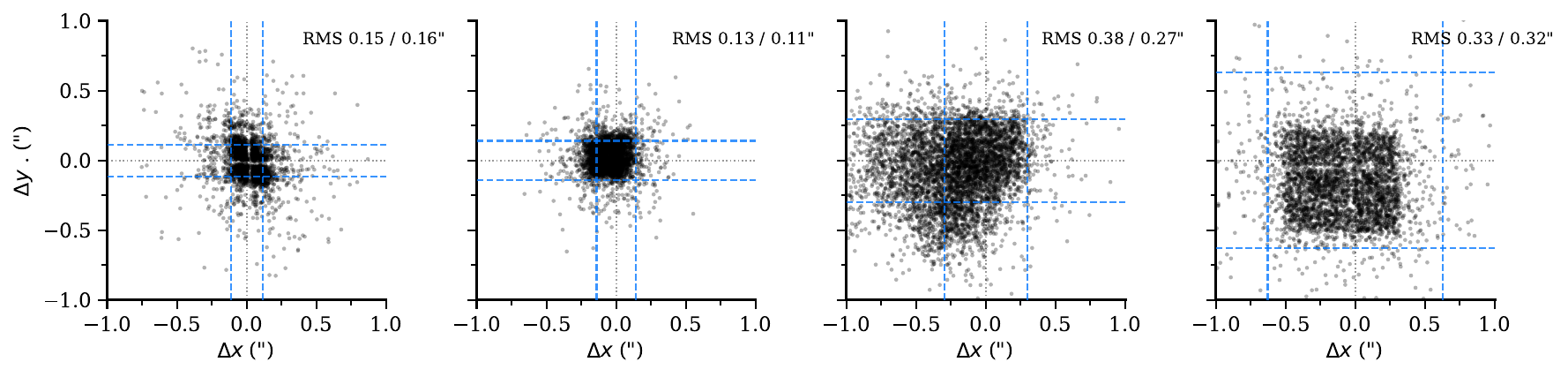}
        \caption{Astrometric residuals: recovered minus injected position, in arcseconds, shown as two-dimensional histograms; dashed lines mark $\pm0.5$ pixels. The RMS per axis is quoted in each panel.}
        \label{fig:res/astrometry}
    \end{subfigure}

    \vspace{0.4em}
    \begin{subfigure}{\textwidth}
        \centering
        \includegraphics[width=\textwidth]{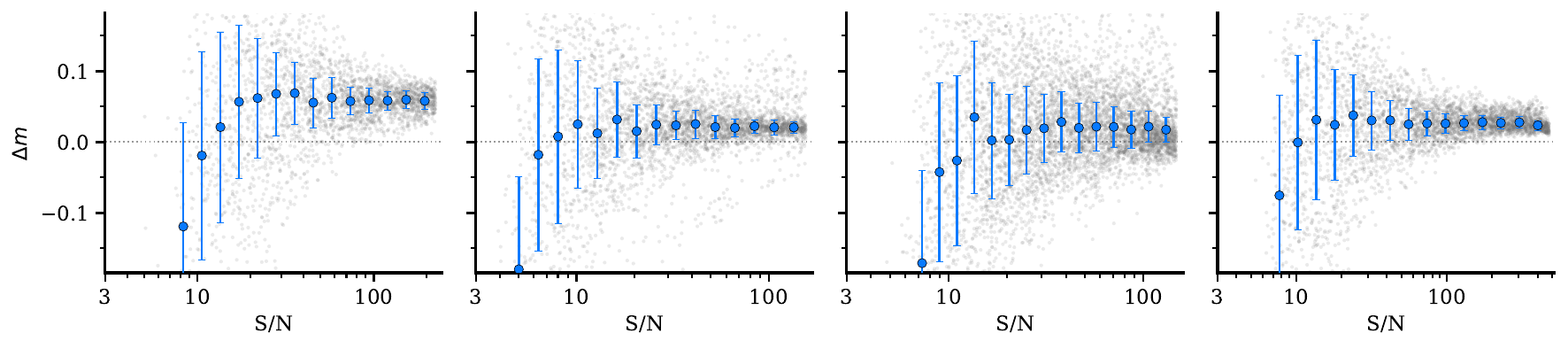}
        \caption{Photometric precision: recovered minus injected magnitude ($\Delta m = m_{\rm measured}-m_{real}$) versus S/N; gray points are individual sources and blue points show the binned median with $\pm1\sigma$ error bars.}
        \label{fig:res/photometry}
    \end{subfigure}

    \caption{Injection--recovery validation of \textsc{GPUPhot} on synthetic images for the four instruments of the survey. Synthetic stars drawn from a realistic brightness distribution are injected onto a flat background with the measured sky level and noise, using the spatially varying PSF model, and recovered by running the full pipeline from scratch.}
    \label{fig:results}
\end{figure*}

Injected magnitudes are drawn uniformly in instrumental magnitude over a range anchored to the real source population, extending from well above the brightest unsaturated stars down to below the noise floor. Sources are placed at random positions subject to a minimum separation of $3\times\mathrm{FWHM}$ from any previously injected star, which keeps blending negligible, and no closer than a border margin to the frame edge. For each instrument we generate several independent realizations, adjusting the number of images and the number of sources per image to the field size, for a total of 10,000 synthetic stars per instrument. Each synthetic frame is written to disk and processed by the complete production pipeline with no knowledge of the injected catalog; the recovered detections are then matched to the truth by position, using a tolerance of $0.5\times\mathrm{FWHM}$. The results are summarized in Figure~\ref{fig:results} and discussed below.

\subsubsection{Detection completeness}
\label{sec:res/completeness}

Completeness is defined as the fraction of injected sources that are recovered, as a function of the predicted signal-to-noise ratio computed from the injected flux and the noise model. Panel~(a) of Figure~\ref{fig:results} shows the completeness (blue) for the four instruments. In all cases the transition is sharp and well described by a logistic function, with a 50\% completeness limit $(\mathrm{S/N})_{50}$ between $5.2$ and $9.3$. Above $(\mathrm{S/N})\gtrsim20$ the completeness plateaus close to unity in every case.

Because a completeness curve alone can always be improved by lowering the detection threshold at the cost of spurious detections, we characterize the false-positive contamination on the same source-free images. Any detection that does not match an injected source is, by construction, a false positive. We quantify this through the precision (red curve; $\mathrm{TP}/(\mathrm{TP}+\mathrm{FP})$), binned by detected S/N. In all instruments the precision rises rapidly and reaches unity for $(\mathrm{S/N})\gtrsim10$: above this level essentially no spurious sources are produced. The contamination is confined to the low-S/N regime near the detection threshold, where it coincides with the incompleteness region, so that the operating range in which both completeness and precision approach unity is broad and clean. The false positives are somewhat more prevalent near threshold for TST, the instrument with the strongest PSF variation. Nonetheless all four instruments reach a pure sample well within the regime where the survey operates.

\subsubsection{Astrometric precision}
\label{sec:res/astrometry}

Panel~(b) of Figure~\ref{fig:results} shows the astrometric residuals in arcseconds between the recovered and the true (injected barycenter) positions. The dashed lines mark $\pm0.5$~pixels at the plate scale of each instrument, so that the sub-pixel nature of the recovery is directly apparent. In all four cases the residual cloud is centered close to the origin and is well contained within the half-pixel lines, with per-axis RMS values ranging from $0.11$ to $0.38\arcsec$. The distributions are compact and approximately isotropic, with only a small residual offset that can be absorbed in the astrometric solution. Some asymmetry is visible for TTT-80cm and ATLAS-Teide as well as TST.

We note that the residuals are quoted in angular units rather than pixels because the instruments differ substantially in plate scale: the undersampled ATLAS camera yields the smallest residual in pixels but not on the sky, and the angular comparison is the physically meaningful one. The larger residuals of TST reflect its wider field and stronger PSF variation, while the well-sampled instruments reach the $\sim\!0.1\arcsec$ level. 

\subsubsection{Photometric precision}
\label{sec:res/photometry}

Panel~(c) of Figure~\ref{fig:results} shows the photometric residuals $\Delta m = m_{\rm measured}-m_{\rm real}$ as a function of S/N; gray points are individual sources and the blue points show the binned median with $\pm1\sigma$ error bars. In all four instruments the median residual is a small, constant offset of $\sim\!+0.05$~mag that remains flat from the completeness limit all the way to the brightest sources. Being independent of brightness and position, it is fully absorbed in the photometric zero point and does not affect relative photometry. Finite growth-curve normalization and pixel-discrete apertures can contribute, although their individual contributions are not isolated here.

The constancy of $\Delta m$ across the entire dynamic range is a key result: it shows that the optimal-aperture scheme, which selects the aperture that maximizes the detection S/N on a source-by-source basis, not only improves the photometric S/N as demonstrated in Figure~\ref{fig:photometric_performance}, but does so without introducing any brightness-dependent bias in the flux measurement. Bright and faint sources receive different apertures, yet the recovered magnitude offset is the same for all of them, confirming that the adaptive aperture and its correction are self-consistent across brightness. The scatter of the band at high S/N sets the floor of the photometric precision, reaching the few-millimagnitude level for the brightest sources and decreases monotonically with brightness, as expected.

\subsection{Comparison with catalogs}
\label{sec:results/catalog}

The synthetic experiment of Section~\ref{sec:results_injection} measures the performance of the estimators against an exact truth catalog, but does not test the calibration of the output onto an absolute reference frame. Two calibration steps are required for that purpose and are applied to every science frame as part of the standard pipeline output: the transformation of the pixel centroids into celestial coordinates and the conversion of the instrumental fluxes into a standard photometric system.

\subsubsection{Astrometric calibration}
\label{sec:results/astrom_cal}

The photometric catalog produced by the pipeline (Section~\ref{sec:photometry}) provides, for every detection, a sub-pixel centroid $\hat{\mu}$ and a detection significance $\nu$. Astrometric calibration begins by selecting from this catalog a clean subset of high-confidence sources: detections above a minimum signal-to-noise threshold are retained, sorted by decreasing significance, and the brightest $N_\star = 500$ are passed to the solver. Restricting the input to the brightest, most reliable centroids both accelerates the asterism matching and minimizes the contamination of the solution by spurious or poorly centroided detections.

The plate solution is obtained with Astrometry.net \citep{Lang2010}, run through a two-tier strategy that favors a fast local solve and resorts to the online service only when necessary. In the first tier the source list is matched against locally stored geometric-hash index files (astrometry.net index series 4100 and 5200, spanning angular scales from a few arcminutes to several degrees) with the reference quad-matching engine. The solve is seeded with two hints derived from the telescope pointing and instrument configuration: a scale hint, bracketing the nominal plate scale and a position hint, restricting the search to a $1\degr$ radius about the commanded field center. When the hinted solve fails---for instance, when the pointing metadata are unreliable---a second, blind local solve is attempted without hints, and only if that also fails is the field submitted to the online Astrometry.net server. In practice the hinted local solve succeeds for the overwhelming majority of science frames and the fallbacks are exercised only for anomalous pointings.

A candidate solution is accepted as soon as the log-odds of the best asterism match exceed a threshold of $100$, which effectively eliminates false positives. Field distortions are captured by fitting a Simple Imaging Polynomial \citep[SIP;][]{Shupe2005} of order three simultaneously with the linear WCS terms, sufficient to describe the optical distortion of the wide-field instruments to well below the centroiding accuracy. The resulting WCS---the linear \texttt{CD} matrix together with the SIP coefficients---is written to the frame header, and the full source catalog is transformed from pixel to celestial coordinates through it, assigning an $(\alpha,\delta)$ pair to every detection.

\begin{figure}[t!]
\centering
\includegraphics[width=\linewidth]{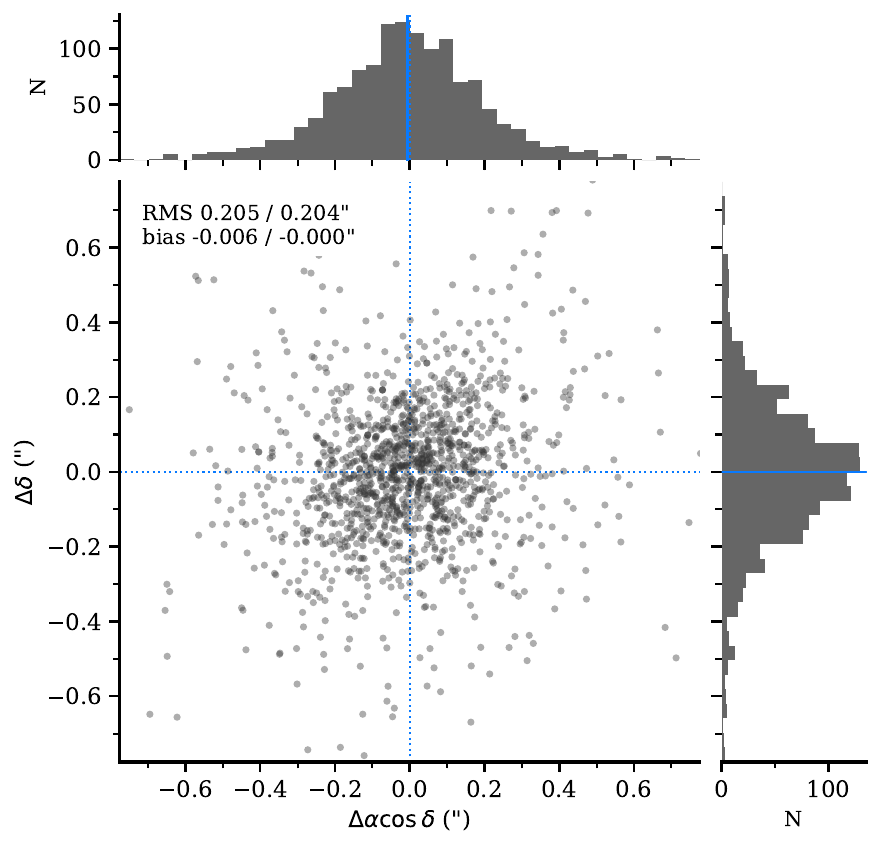}
\caption{Astrometric accuracy of the \textsc{GPUPhot} solution on a representative science frame. \textit{Main panel:} residuals between the calibrated source positions and their cross-matched counterparts in the reference catalog, in the sense recovered minus catalog, decomposed into $\Delta\alpha\cos\delta$ and $\Delta\delta$ in arcseconds. \textit{Top and right panels:} marginal histograms of the two residual components. The per-axis RMS and bias are quoted in the main panel and the dotted lines mark the origin. The distribution is compact, approximately isotropic and centered on zero, with a per-axis RMS below the pixel scale of this frame ($0.283 ^{\prime\prime}$px$^{-1}$) and no measurable systematic offset.
\label{fig:astrometry_image}}
\end{figure}

The accuracy of the solution is assessed by cross-matching the calibrated source positions against the reference catalog and examining the residuals. Figure~\ref{fig:astrometry_image} shows the two-dimensional distribution of the offsets $\Delta\alpha\cos\delta$ and $\Delta\delta$ for a representative field taken with one of the TTT-80cm, together with their marginal histograms. The residual cloud is tightly concentrated about the origin, with per-axis RMS values of $\sim0.2^{\prime\prime}$ and negligible systematic biases in R.A. and Dec., respectively. The absence of any measurable bias confirms that the SIP solution absorbs the field distortion without leaving a coherent offset, while the RMS---a fraction of a pixel at the plate scale of this instrument---sets the astrometric precision of the survey and is consistent with the sub-pixel performance measured in the controlled experiment of Section~\ref{sec:res/astrometry}.

\subsubsection{Photometric calibration}
\label{sec:results/photom_cal}

With the WCS in place, the instrumental fluxes are tied to a standard system through a photometric zero point (ZP). \textsc{GPUPhot} first queries, via the VizieR service \citep{Ochsenbein2000}, a photometric reference catalog over a cone of radius equal to half the field diagonal centered on the frame. The catalog is selected automatically from the observing filter and the field declination: Pan-STARRS\,DR1 \citep{Chambers2016} for the SDSS-like broad-band filters over the accessible sky, Gaia\,DR3 \citep{Gaia2023} for the unfiltered configuration and the SkyMapper Southern Survey \citep{Onken2024} for fields south of $\delta=-30\degr$ beyond the Pan-STARRS footprint. For each reference star the query returns its celestial coordinates, its magnitude in the band matched to the observation and a color index defined below.

The calibrated sources are cross-matched against the reference catalog by position, using a $k$-d tree with a matching tolerance of $1.5\times\mathrm{FWHM}$. For each matched star the instrumental magnitude is formed from the calibrated flux $S$ of Section~\ref{sec:photometry}, normalized by the exposure time $t_{\rm exp}$,
\begin{equation}
    m_{\rm inst} = -2.5\log_{10}\!\left(\frac{S}{t_{\rm exp}}\right),
    \label{eq:m_inst}
\end{equation}
and the zero point is defined by the relation $m_{\rm cat} = m_{\rm inst} + \mathrm{ZP}$, i.e.\ as the offset between the catalog and instrumental magnitudes at unit slope.

\begin{figure}[t!]
\centering
\includegraphics[width=\linewidth]{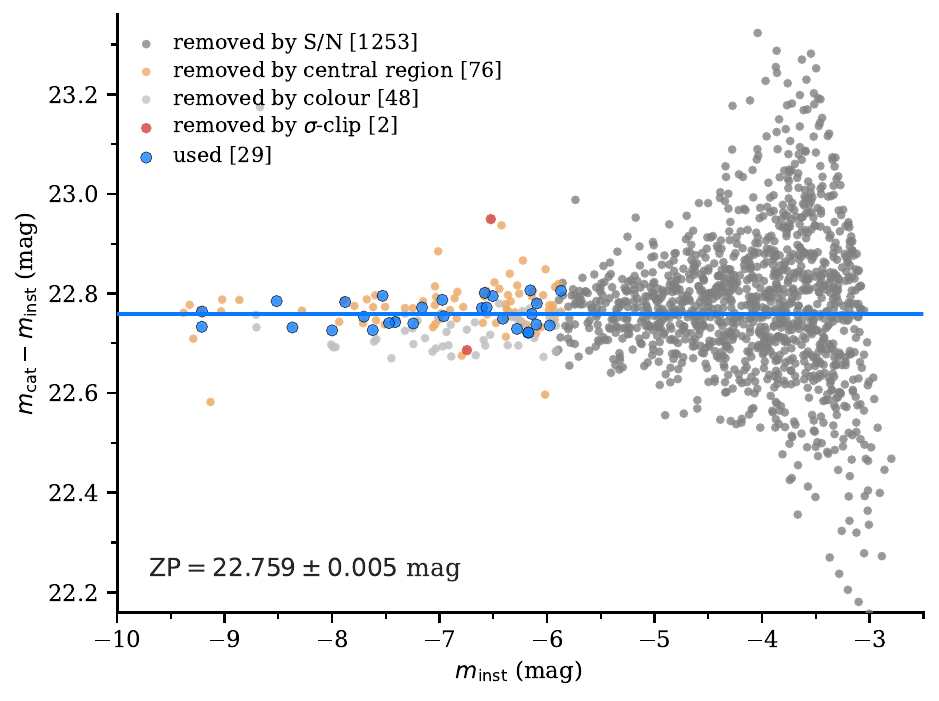}
\caption{Photometric zero-point determination on a representative science frame. The difference between the catalog and instrumental magnitudes, $m_{\rm cat}-m_{\rm inst}$, is shown as a function of instrumental magnitude for all cross-matched stars. Points are color-coded by the selection stage that discarded them---signal-to-noise (gray), central region (orange), color (light gray) and the final $\sigma$-clip (red)---while the blue points are the $N=29$ stars retained for the fit. The horizontal blue line marks the adopted zero point, $\mathrm{ZP} = 22.759 \pm 0.005$~mag, estimated as the sigma-clipped mean of the retained offsets at fixed unit slope. The legend gives the number of stars in each category.
\label{fig:zeropoint}}
\end{figure}

Before the zero point is fitted, the matched sample is filtered by three criteria designed to isolate a clean, well-behaved set of calibrators, illustrated by the color coding in Figure~\ref{fig:zeropoint}. First, a color cut retains only stars of near-solar color, $|\mathcal{C}| < \Delta\mathcal{C}$ with $\Delta\mathcal{C}=0.6$~mag by default, where $\mathcal{C} \equiv (B-V) - (B-V)_\odot$ is the deviation from the solar color $(B-V)_\odot = 0.65$. The index $\mathcal{C}$ is computed from the catalog photometry through published transformations: the $g,r\to B,V$ relations of \citet{Tonry2012} for Pan-STARRS and SkyMapper, and a quadratic function of the $B_P-R_P$ color for Gaia \citep{Riello2021}. This particular choice reflects the origin of \textsc{GPUPhot}, developed to process the large volume of Solar System observations produced by the TTT, TST and ATLAS-Teide surveys, whose targets have approximately solar colors and for which a calibration anchored on solar-type stars is therefore the natural one. The criterion is not intrinsic to the
method: the width $\Delta\mathcal{C}$ is a free parameter, the reference color can be set to any value appropriate to the science case and the cut can be disabled altogether---in which case the full color range of the matched sample enters the fit, at the cost of a zero point that absorbs the unfitted color term of the sample as a whole. Second, a signal-to-noise cut, $30 < \mathrm{S/N} < 300$, excludes both faint stars, whose photometry is noise-dominated, and the brightest stars, which may approach the non-linear or saturated regime. Third, a central-region cut keeps only sources within the central portion of the detector, where the PSF is most symmetric and vignetting is negligible.

The zero point is then estimated as a robust, sigma-clipped mean of the per-star offsets $y_i = m_{{\rm cat},i} - m_{{\rm inst},i}$, holding the slope fixed at unity. Starting from the full surviving sample, the median $\tilde{y}$ and standard deviation $\sigma_y$ are computed, and stars deviating from the median by more than $k\sigma_y$ (with $k=2.5$) are rejected; the procedure iterates until the retained set is stable or drops below three stars, and never exceeds ten iterations. The zero point and its uncertainty are the mean and standard error of the surviving offsets.

Although a free slope can absorb residual color or non-linearity trends within a single frame, we find empirically---on time series of repeated observations of the same field---that the pure offset yields a markedly more stable zero point from image to image, i.e. a lower scatter in $\mathrm{ZP}(t)$, because it does not couple the calibration to frame-to-frame changes in the calibrator distribution. The color and S/N cuts above already remove the trends that a free slope would otherwise be tuned to capture.

Figure~\ref{fig:zeropoint} shows the outcome for the same representative field. The offset $m_{\rm cat}-m_{\rm inst}$ is plotted against the instrumental magnitude for every matched star, color-coded by the criterion that removed it with the $N=29$ retained calibrators highlighted and the fitted zero point drawn as a horizontal line. The retained points scatter tightly about a constant offset with no discernible trend in magnitude, confirming that the unit-slope assumption holds over the calibrated range.

\section{GPU implementation}
\label{sec:gpu}

Most of the algorithms presented in Sections~\ref{sec:defects}--\ref{sec:photometry} are not new in their purpose, and several have been used before in whole or in part. What is deliberate here is their reformulation: each has been recast so that it reduces to a convolution with a position-independent kernel or to an element-wise operation on full-frame maps. The justification for that redesign is not the individual algorithm, for which a conventional formulation would often be simpler, but the requirement that the complete pipeline run on a GPU over large numbers of large images. A Graphics Processing Unit applies a single instruction to many data elements concurrently and reaches its nominal throughput only for operations that are uniform across those elements. Under this constraint no stage iterates over individual sources, and no stage has a cost that depends on the number of distinct PSFs present in the field: the position dependence is carried by coefficient maps computed once per frame and applied as element-wise factors. The per-frame cost is set by the number of transforms rather than by the source count, and its distribution across stages follows from the same structure. Constructing the eigenmode basis and interpolating its coefficient maps across the focal plane together account for a small fraction of the per-frame wall-clock time, comparable to that of background estimation, while the per-source growth curves of Equation~(\ref{eq:aperture_flux_fft}), evaluated at every radius of the grid of Equation~(\ref{eq:aperture_radii_set}), dominate in fields of high source density. Describing the PSF empirically as a function of position is therefore not appreciably more expensive than assuming it to be constant.

The algorithms are implemented in \textsc{GPUPhot}, within a framework that handles image and data ingestion and output, the management of computational resources, access to external catalogs and the storage of point-source measurements in a single environment. On a consumer-grade NVIDIA RTX~3060, a $2\rm{k}\times2\rm{k}$ CCD frame is reduced in under $7$\,s and a $6\rm{k}\times6\rm{k}$ sCMOS frame of moderate source density in under $25$\,s, in both cases from the raw frame to the calibrated catalog: defect correction, background and noise-map estimation, PSF modeling, detection, photometry and astrometric and photometric calibration. On an NVIDIA H100~PCIe, a datacenter accelerator intended for high-performance computing workloads, the same frames require approximately $5$\,s and $8$\,s, and a $14\rm{k}\times11\rm{k}$ (151-Mpx) wide-field image containing more than $10^{4}$ sources is reduced in under $40$\,s. These are production timings rather than a controlled CPU-to-GPU comparison. Because the memory footprint of a frame of this size leaves headroom on an $80$\,GB device, at least four reductions run concurrently on a single accelerator, bringing the effective throughput to better than one wide-field image every $10$\,s. 

\textsc{GPUPhot} has been in routine operation at the Teide Observatory since February 2023, when it began reducing the data stream of the TTT and TST, and since January 2025 for ATLAS-Teide. It has to date reduced 3.7 million frames, amounting to $1.2$~PB of imaging, at a sustained average of $\sim\!5000$ frames and $1.5$~TB per night across all telescopes. A companion paper \citep{LemesPerera2026} describes the implementation of the framework and quantifies the execution time of each pipeline stage across image sizes, source densities and a range of GPU platforms, together with the scripts required to reproduce those measurements. It also compares against the CPU tools \texttt{sep} and Photutils, both for the overall tools and for detection on the same input arrays. The former comparison has different pipeline scopes, and the latter still returns different source lists; neither is an identical-output end-to-end speed-up. A fully equivalent benchmark would require matching and optimizing the individual stages, including their accuracy criteria and I/O.  The code is publicly available in \url{https://github.com/Light-Bridges/GPUPhot}.

\section{Conclusions}
\label{sec:conclusions}

We have described the algorithms that make up \textsc{GPUPhot}, covering the analysis of a calibrated frame through to the final catalog: correction of cosmic rays and detector defects, background and noise-map estimation, the construction of an empirical PSF model that varies across the focal plane, matched-filter source detection and adaptive aperture photometry with its empirical correction and noise model. Each is presented as a self-contained procedure, and several are applicable beyond the photometric pipeline in which they are assembled here.

We validated the pipeline through a controlled injection--recovery experiment on four instruments spanning an order of magnitude in plate scale, from the small, well-sampled TTT-2m CCD field to the severely undersampled ATLAS-Teide sCMOS wide field. The detection is complete and pure over a broad operating range, with $50\%$ completeness limits between $(\mathrm{S/N})_{50}\simeq5$ and $9$ and a precision that reaches unity above $\mathrm{S/N}\simeq10$. The astrometry is sub-pixel in every case, with per-axis residuals of $0.11$--$0.38\arcsec$. The photometry shows a constant offset of $\sim0.05$~mag, independent of brightness and position, that is fully absorbed into the zero point, reaching a few-millimagnitude floor for the brightest sources. The agreement with external catalogs confirms this performance on real data: the astrometric solution reaches a sub-pixel RMS with no measurable bias and the photometric zero point is recovered with millimagnitude precision.

Taken together, these results show that a consistently kernel-based formulation makes the reduction of image archives at the petabyte scale tractable at low computational cost, without sacrificing completeness or astrometric and photometric accuracy.

\begin{acknowledgments}
This research made use of storage and computing capacity at ASTROPOC’s EDGE computing center in Tenerife in the form of Indefeasible Computer Rights (ICR), provided by Light Bridges in cooperation with Bechtle and LENOVO. Additionally, this work is based on observations made with the Two-meter Twin Telescope (TTT) and the Transient Survey Telescope (TST), located at the Teide Observatory of the Instituto de Astrof\'{\i}sica de Canarias (IAC) and operated by Light Bridges in Tenerife, Canary Islands, Spain, using observation time rights (DTO). Dr Antonio Maudes’s economic and legal insights were instrumental in shaping this work. M.R.A. and J.L. We acknowledge financial support through grant PID2024-160618NB-C22, “Hydrated Minerals and Organic Compounds in Primitive Asteroids 2”, funded by MICIU/AEI/10.13039/501100011033 and by ERDF, EU, and through the Severo Ochoa Centre of Excellence accreditation awarded to the Instituto de Astrof\'{\i}sica de Canarias, grant CEX2025-001609-S, funded by MICIU.

\end{acknowledgments}

\begin{contribution}

MRA conceptualized the research, designed the core algorithms and software and wrote the original draft of the manuscript. SLP optimized the code and validated the computational results. MSR and JL provided supervision and contributed to methodological discussions. All authors reviewed and edited the manuscript and approved its final version.

\end{contribution}

\facilities{TTT (FERVOR, COLORS), TST (FERVOR), ATLAS-Teide.}

\newpage
\appendix

\section{Statistical derivation of the detection statistic}
\label{app:detection}

This appendix derives in detail the maximum-likelihood detection statistic introduced in Section~\ref{sec:psf/isolated}. We carry out the derivation for a general point spread function $P$ and specialise to the fixed Gaussian kernel $G$ used for isolated-star detection only at the end. 

\subsection{Image model and assumptions}
\label{app:detection/model}

We take as our starting point the isolated point-source model of Equation~(\ref{eq:point_source_model_general}),
\begin{equation}
    M_x = \alpha P_{x-\mu} + \epsilon_x,
    \label{eq:app_model}
\end{equation}
where $M_x$ is the observed (background-subtracted) value at pixel $x$, $\alpha$ is the source flux, $P$ is the PSF normalized to unit sum ($\sum_x P_x = 1$) and centered at the trial position $\mu$, and $\epsilon_x$ is the noise at pixel $x$. We adopt three standard assumptions on the noise $\epsilon_x$:
\begin{enumerate}
    \item \textbf{Gaussianity:} $\epsilon_x \sim \mathcal{N}(0, \sigma_x^2)$, i.e. each $\epsilon_x$ follows a Gaussian distribution of zero mean and variance $\sigma_x^2$ given by Equation~(\ref{eq:rms}). 
    \item \textbf{Pixel independence:} the noise values at different pixels are statistically uncorrelated, $\mathrm{Cov}[\epsilon_x, \epsilon_{x'}] = 0$ for $x \neq x'$. This is a reasonable approximation for typical astronomical detectors, where pixel-to-pixel correlations introduced by charge diffusion or readout electronics are negligible relative to the dominant noise sources.
    \item \textbf{Background-dominated regime:} the source Poisson term in Equation~(\ref{eq:total_noise}) is negligible compared to the sky and read-noise contributions, so that $\mathrm{Var}[\epsilon_x] \approx \sigma_x^2$. This regime is the natural one for source detection: the null hypothesis is that no source is present, so the source Poisson term does not enter the noise budget.
\end{enumerate}

\subsection{The likelihood function}
\label{app:detection/likelihood}

Under the Gaussianity assumption, the probability density of observing the value $M_x$ at pixel $x$, given the model parameters $(\alpha, \mu)$, is
\begin{equation}
    p(M_x \mid \alpha, \mu) = \frac{1}{\sqrt{2\pi\sigma_x^2}}\exp\!\left[-\frac{(M_x - \alpha P_{x-\mu})^2}{2\sigma_x^2}\right].
\end{equation}
Pixel independence implies that the joint probability of observing the entire image factorises as
\begin{equation}
    \mathcal{L}(\alpha, \mu) \equiv p(M \mid \alpha, \mu) = \prod_x p(M_x \mid \alpha, \mu).
    \label{eq:app_likelihood}
\end{equation}
Taking the logarithm converts the product into a sum, while the per-pixel normalization factors collect into a single additive term $C \equiv -\frac{1}{2}\sum_x \ln(2\pi\sigma_x^2)$:
\begin{equation}
    \ln \mathcal{L}(\alpha, \mu) = -\frac{1}{2}\sum_x \frac{(M_x - \alpha P_{x-\mu})^2}{\sigma_x^2} + C.
    \label{eq:app_loglikelihood}
\end{equation}
Maximising this log-likelihood is equivalent to minimising the inverse-variance--weighted sum of squared residuals, i.e. a weighted $\chi^2$.

\subsection{Maximum-likelihood estimator}
\label{app:detection/mle}

At fixed trial position $\mu$, the maximum-likelihood estimate of the flux $\hat{\alpha}(\mu)$ is obtained by setting the derivative of $\ln \mathcal{L}$ with respect to $\alpha$ to zero. Because the background-dominated assumption makes $\sigma_x$ independent of $\alpha$, the noise variance does not contribute to the derivative, so this yields a single linear equation:
\begin{equation}
    \left.\frac{\partial \ln \mathcal{L}}{\partial \alpha}\right|_{\hat{\alpha}} = \sum_x \frac{\left(M_x - \hat{\alpha} P_{x-\mu}\right)P_{x-\mu}}{\sigma_x^2} = 0.
\end{equation}
Rearranging,
\begin{equation}
    \hat{\alpha}(\mu)\sum_x \frac{P_{x-\mu}^2}{\sigma_x^2} = \sum_x \frac{M_x P_{x-\mu}}{\sigma_x^2},
\end{equation}
which yields
\begin{equation}
    \hat{\alpha}(\mu) = \frac{\displaystyle\sum_x M_x P_{x-\mu}/\sigma_x^2}{\displaystyle\sum_x P_{x-\mu}^2/\sigma_x^2}.
    \label{eq:app_alpha_hat}
\end{equation}
With $P$ replaced by the Gaussian kernel $G$, Equation~(\ref{eq:app_alpha_hat}) is equivalent to Equation~(\ref{eq:alpha_hat}).

\subsection{Variance of the estimator}
\label{app:detection/variance}

The weights in Equation~(\ref{eq:app_alpha_hat}) depend only on the kernel and the noise map, not on the data themselves, so $\hat{\alpha}(\mu)$ can be written as a linear combination of the image pixels:
\begin{equation}
    \hat{\alpha}(\mu) = \sum_x w_x M_x, \qquad w_x \equiv \frac{P_{x-\mu}/\sigma_x^2}{\displaystyle\sum_{x'} P_{x'-\mu}^2/\sigma_{x'}^2}.
\end{equation}
For a linear combination of independent random variables, the variance is $\mathrm{Var}[\hat{\alpha}(\mu)] = \sum_x w_x^2 \mathrm{Var}[M_x]$. Under the background-dominated assumption, $\mathrm{Var}[M_x] = \mathrm{Var}[\epsilon_x] = \sigma_x^2$, and letting $K \equiv \sum_x P_{x-\mu}^2/\sigma_x^2$ denote the denominator,
\begin{equation}
    \mathrm{Var}\!\left[\hat{\alpha}(\mu)\right] = \frac{1}{K^2}\sum_x \frac{P_{x-\mu}^2}{\sigma_x^4} \sigma_x^2 = \frac{1}{K^2}\sum_x \frac{P_{x-\mu}^2}{\sigma_x^2} = \frac{K}{K^2} = \frac{1}{K}.
\end{equation}
Therefore
\begin{equation}
    \sigma_{\hat{\alpha}}^2(\mu) = \left[\sum_x P_{x-\mu}^2/\sigma_x^2\right]^{-1}.
    \label{eq:app_variance}
\end{equation}
Physically, this weighting extracts all the information about the flux that the image contains: it matches the filter to the shape of the source through $P_{x-\mu}$, giving most weight to the pixels where the PSF deposits the most light, and to the reliability of each pixel through $1/\sigma_x^2$, down-weighting those dominated by noise. No alternative way of weighting and combining the pixels yields a lower-variance unbiased estimate of the flux, and the resulting matched-filter form is the reason it is the method of choice for point-source detection.

\subsection{Locally stationary approximation}
\label{app:detection/local}

The exact detection statistic $\nu_P(\mu) = \hat{\alpha}(\mu)/\sigma_{\hat{\alpha}}(\mu)$ carries the position-dependent weights $1/\sigma_x^2$ inside the sums. These weights break the convolution structure and, therefore, preclude evaluation by FFT-based convolution. We thus adopt the locally stationary approximation: if $\sigma_x$ varies on spatial scales much larger than the kernel size, then within a region of radius comparable to the kernel width centered on $\mu$, $\sigma_x \approx \sigma_\mu$.
Substituting this into Equation~(\ref{eq:app_alpha_hat}) factors $\sigma_\mu^{-2}$ out of both sums, yielding
\begin{equation}
    \hat{\alpha}(\mu) \approx \frac{\sum_x M_x P_{x-\mu}}{\sum_x P_{x-\mu}^2} = \frac{(M * \bar{P})_\mu}{A_P},
    \label{eq:app_alpha_local}
\end{equation}
where $\bar{P}_x \equiv P_{-x}$ is the reflected kernel, $(M*\bar{P})_\mu \equiv \sum_x M_x \bar{P}_{\mu-x} = \sum_x M_x P_{x-\mu}$ is the cross-correlation evaluated at $\mu$, and $A_P \equiv \sum_x P_x^2$ is the effective kernel area, independent of $\mu$. Analogously, the estimator variance becomes
\begin{equation}
    \sigma_{\hat{\alpha}}^2(\mu) \approx \frac{\sigma_\mu^2}{A_P}.
    \label{eq:app_variance_local}
\end{equation}
Combining Equations~(\ref{eq:app_alpha_local}) and~(\ref{eq:app_variance_local}),
\begin{equation}
    \nu_P(\mu) = \frac{\hat{\alpha}(\mu)}{\sigma_{\hat{\alpha}}(\mu)} = \frac{(M*\bar{P})_\mu/A_P}{\sigma_\mu/\sqrt{A_P}} = \frac{(M * \bar{P})_\mu}{\sigma_\mu \sqrt{A_P}}.
    \label{eq:app_nu_final}
\end{equation}
For isolated-star detection the PSF is not yet known, and we replace $P$ by the fixed Gaussian kernel $G$ ($A_P \to A_G$, $\bar{P} \to \bar{G}$), recovering Equation~(\ref{eq:conv_sigma_isolated}) of the main text. This result is equivalent to Equation~(10) of \citet{Bosch2018}.

\bibliography{sample7}{}
\bibliographystyle{aasjournal}

\end{document}